\documentclass[superscriptaddress,twocolumn,noshowpacs]{revtex4}
\usepackage{graphicx}
\usepackage{textcomp}
\usepackage{amsmath}
\usepackage{commath}
\usepackage{color}

\begin{document}
	
	\title{Phase transitions in 2D multistable mechanical metamaterials via collisions of soliton-like pulses}
	
	\author{Weijian Jiao}
	\affiliation{Department of Mechanical Engineering and Applied Mechanics, University of Pennsylvania, Philadelphia, Pennsylvania 19104, USA}
	
	\author{Hang Shu}
	\affiliation{Department of Mechanical Engineering and Applied Mechanics, University of Pennsylvania, Philadelphia, Pennsylvania 19104, USA}
	
	\author{Vincent Tournat}
	\affiliation{Laboratoire d'Acoustique de l'Université du Mans (LAUM), UMR 6613, Institut d'Acoustique - Graduate School (IA-GS), CNRS, Le Mans Université, France }
	
	\author{Hiromi Yasuda}
	\affiliation{Department of Mechanical Engineering and Applied Mechanics, University of Pennsylvania, Philadelphia, Pennsylvania 19104, USA}
	\affiliation{Aviation Technology Directorate, Japan Aerospace Exploration Agency, Mitaka, Tokyo 1810015, Japan}
	
	\author{Jordan R. Raney}
	\email{raney@seas.upenn.edu}
	\affiliation{Department of Mechanical Engineering and Applied Mechanics, University of Pennsylvania, Philadelphia, Pennsylvania 19104, USA}
	
	\begin{abstract}
		In this work, we report observations of phase transitions in 2D multistable mechanical metamaterials that are initiated by collisions of soliton-like pulses in the metamaterial. Analogous to first-order phase transitions in crystalline solids, we experimentally and numerically observe that the multistable metamaterials support phase transitions if the new phase meets or exceeds a critical nucleus size. If this criterion is met, the new phase subsequently propagates in the form of transition waves, converting the rest of the metamaterial to the new phase. More interestingly, we observe that the critical nucleus can be formed via collisions of soliton-like pulses. Moreover, the rich direction-dependent behavior of the nonlinear pulses enables 
		control of the location of nucleation and the spatio-temporal shape of the growing phase.
	\end{abstract}
	
	\maketitle
	
	\begin{figure*}[htbp]
		\centerline{ 
			\includegraphics[width=1\textwidth]{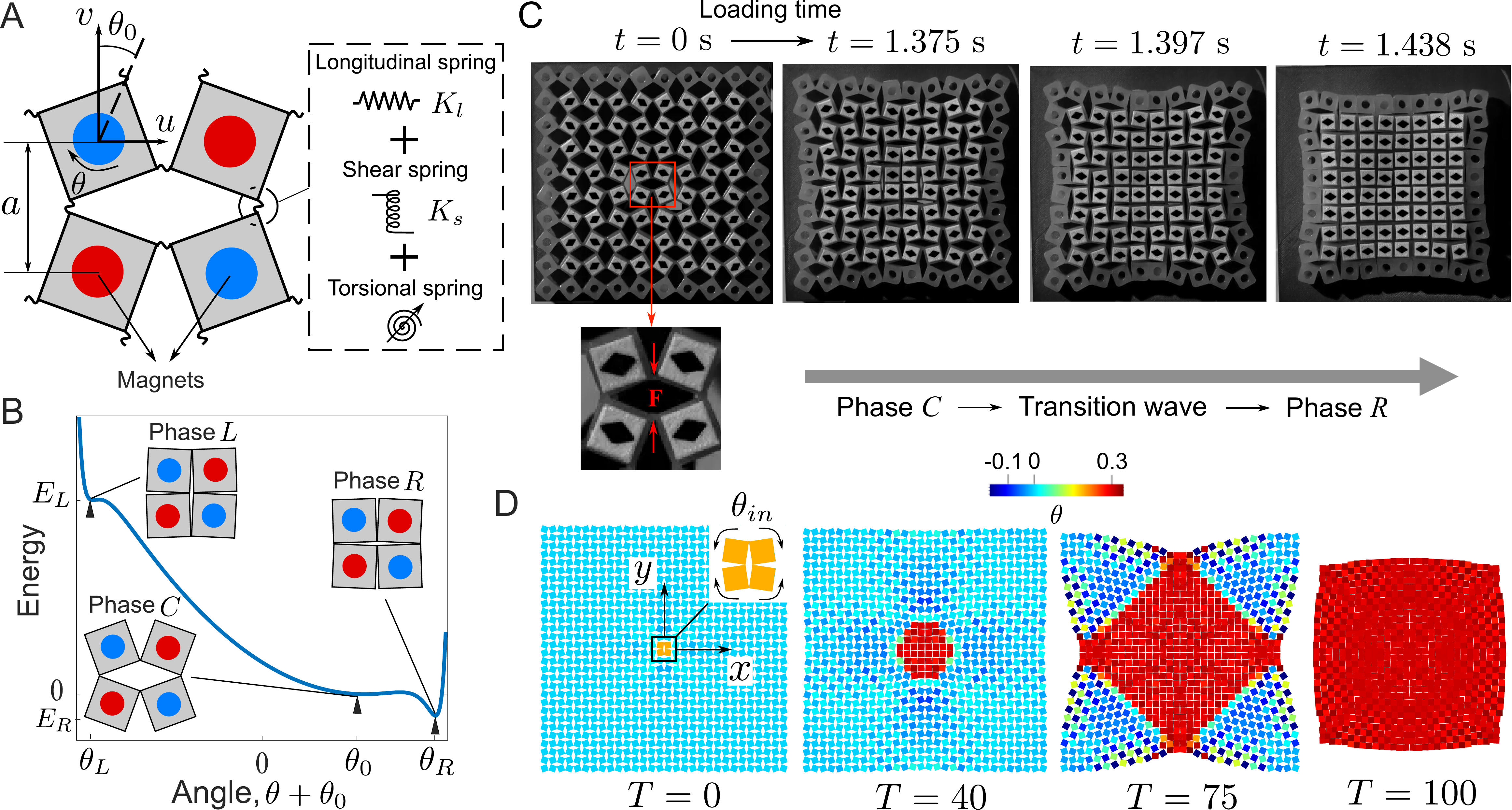}}
		\caption{(A) Schematic of a four-square building block of the metamaterial and (B) its multistable potential energy landscape. (C)~Optical snapshots of an experimental specimen with the center four squares subjected to quasistatic rotation via the application of the force $F$; 
			this causes the formation of a new phase and 
			its eventual growth outward through the rest of the metamaterial. (D)~Snapshots of quasistatic nucleation and growth observed via numerical simulation 
			for a system comprising $30\times30$ squares. The positive rotational direction is defined in a way that rotates the squares from the initial Phase~\textit{C} to the new Phase~\textit{R}.  } 
		\label{fig:Experimental_results}
	\end{figure*}
	
	Nonlinear mechanical metamaterials have received significant attention in the past decade, due to their versatile static and dynamic behavior\cite{Florijn_PRL_2014,bertoldi2017flexible,Deng_2021}, and the ability to tune their response. For example, nonlinear mechanical metamaterials have been previously designed that exhibit tunable stiffness~\cite{Florijn_PRL_2014}, Poisson’s ratio~\cite{Chen_PRApplied_2017}, thermal expansion~\cite{Wang_PRL_2016}, and band gaps~\cite{Wang_PRL_2014,Shan_2014}. 
	Nonlinear mechanical metamaterials often exhibit 
	rich amplitude-dependent properties, such as weakly nonlinear harmonic waves~\cite{cabaret2012amplitude, Jiao_prapplied_2018, Jiao_pre_2021}, cnoidal waves~\cite{Mo2019}, solitary waves~\cite{Deng2017,Deng_2018}, and transition waves~\cite{Nadkarni2016,Raney2016,Hwang2018,Jin2020}.  
	
	One particular class of mechanical metamaterial obtains its nonlinear properties from the rotation of periodic internal features, such as squares connected at their hinges. Systems based on the rotating-squares mechanism have long been studied due to their interesting static properties (i.e., their auxetic characteristicis)~\cite{Grima2000,Grima2013}. More recently, it has been observed that they are also capable of propagating a variety of nonlinear waves~\cite{Deng2017,Mo2019,Yasuda2020}. A notable example is the propagation of vector solitons, which have coupled translational and rotational degrees of freedom (DOFs) and can display distinct solitary modes for different propagation directions~\cite{Deng2019}. 
	Interactions of these nonlinear waves have also been investigated, albeit mostly for one-dimensional systems.
	Due to the coupling between different DOFs, which is less often considered in Hertzian granular media~\cite{nesterenko1984propagation, Coste1997,Daraio_2006,SEN200821,Shen_pre2014}, 
	the collision of vector solitons has been shown to exhibit anomalous phenomena, including repelling, destruction, etc., in addition to classical soliton collisions~\cite{Deng2019_collision}. 
	
	Recently, the dynamics of multistable versions of these systems have also been studied. For example, multistability can be achieved by introducing permanent magnets~\cite{Yasuda2020,Korpas2021}, which produces multiple energy minima, each associated with equilibrium angles that the squares can snap between. If squares are rotated from one stable angle to another, it is possible for this reconfiguration to propagate throughout the structure in the form of a transition wave. In addition, the collision of transition waves of incompatible type can cause formation of stationary domain walls, which can be exploited for the design of reconfigurable metamaterials~\cite{Yasuda2020}. 
	
	Here, we investigate collisions of nonlinear, soliton-like pulses in 2D multistable systems of rotating squares, and how these collisions can be used to remotely nucleate phase transitions at arbitrary locations. As a first step, we experimentally and numerically show how phase transitions can be initiated via quasistatic rotation of a ``critical nucleus'' of squares, analogous to nucleation during first-order phase transitions 
	\cite{porter2009phase,james1986displacive}. Note, in this work, the phase transitions are enabled by multistability, which is achieved by embedding magnets in the squares. This is in contrast with other work~\cite{YANG2016,Deng_PNAS_2020,Bossart_PNAS_2021}, in which phase transitions are induced by applying static precompression to the entire system, or by dynamic recoil~\cite{Liang_pnas_2022}. Second, we investigate the criteria necessary for collisions of soliton-like pulses to induce this phase transition. Finally, we describe how the anisotropy associated with the symmetry of the system produces direction-dependent nucleation and propagation of the phase transition. 
	These fundamental behaviors could enable new insights for the design of reconfigurable, shape-transforming, and deployable mechanical metamaterials.
	
	\section*{Phase transitions in multistable metamaterials}
	We start by experimentally and analytically characterizing the energy landscape of the building block of the mechanical system, i.e., a $2\times2$ set of squares. 
	To experimentally measure the behavior of such a system, we fabricate an elastomeric building block, following a conventional molding-casting process. 
	Specifically, we design and 3D print a mold (MakerGear M2, polylactic acid). We then pour silicone precursor (Dragon Skin 10) into the mold and allow it to cure. The squares have side length 12~mm and are connected by thin hinges of thickness 1.5~mm. Permanent magnets are inserted into each square (SI Appendix, Section~1 for fabrication details).
	A schematic of the building block is shown in Fig.~\ref{fig:Experimental_results}A. The competition between the strain energy of the hinge and the interaction of the magnets gives the squares three stable angles~\cite{Yasuda2020}. Each of these corresponds to a local minimum in the potential energy landscape (Fig.~\ref{fig:Experimental_results}B). Then, in order to quantify the effects of different design parameters, we introduce a discrete model capable of capturing 
	the multistable energy landscape. Each square, assumed to be a rigid body with mass $M$ and moment of inertia $J$, has two translational degrees of freedom ($u$ and $v$) and one rotational degree of freedom ($\theta$). Each hinge is modeled by three springs (Fig.~\ref{fig:Experimental_results}A): a linear longitudinal spring with stiffness $K_l$, a linear shear spring with stiffness $K_s$, and a nonlinear torsional spring with potential energy $E_\theta(\Delta \theta)$ expressed as
	\begin{widetext}
		\begin{align}\label{Potential_theta}
			E_\theta(\Delta \theta)&=\frac{1}{2}K_\theta(\Delta \theta)^2 +V_{\mathrm{Morse}}(\Delta \theta) , \\
			V_{\mathrm{Morse}}(\Delta \theta)
			&=A\left[e^{2\alpha(\Delta \theta +2\theta_0-2\theta_M)}-2e^{\alpha(\Delta \theta +2\theta_0-2\theta_M)}\right]\nonumber\\
			&+A\left[e^{-2\alpha(\Delta \theta +2\theta_0+2\theta_M)}-2e^{-\alpha(\Delta \theta +2\theta_0+2\theta_M)}\right], \label{Potential_theta_M}
		\end{align}
	\end{widetext}
	where $K_\theta$ is the linear torsional spring constant, $\theta_0$ is the initial equilibrium angle, $\Delta \theta$ is the relative angle of the hinge, and $V_{\mathrm{Morse}}$ is the Morse potential, which is used to empirically describe the nonlinear magnetic interactions between squares. In Eq.~\ref{Potential_theta_M}, $A$ and $\alpha$ define the depth and width of the Morse potential, respectively, and $\theta_M$ determines the equilibrium points. 
	To obtain these parameters for the numerical simulations, we conduct experimental tensile tests using a commercial quasistatic test system (Instron model 68SC-5) with custom fixtures  (SI Appendix, Fig.~S2 and S3). These are designed to allow the squares to rotate during the tests. Then, Eq.~\ref{Potential_theta} gives the energy landscape of the building block, which exhibits three distinct phases (labeled as Phase \textit{L}, \textit{C}, and \textit{R}), as shown in the inset of Fig.~\ref{fig:Experimental_results}B.
	
	Before considering whether collisions of impulses can induce a phase transition in the system, we first seek to understand the threshold for nucleation of a phase transition more generally. We assume that the system is initially in Phase \textit{C}, and that a small number of squares are forced to rotate to the new Phase \textit{R}; we then experimentally and numerically observe whether this forced rotation nucleates a new phase, which can propagate throughout the rest of the structure. To confirm this experimentally, we fabricate a larger prototype of size $10\times10$ squares, following the same procedures described earlier (note, to reduce the effect of the boundaries on the behavior of the mechanical system, magnets are not embedded in the exterior squares). Nucleation is induced by quasistatically forcing a $2\times2$ building block at the center of the specimen to undergo the transition. 
	The entire specimen is observed to subsequently undergo a phase transition, as shown in the optical images of Fig.~\ref{fig:Experimental_results}C, obtained via a high-speed camera (Photron FASTCAM Mini AX; Movie S1 and SI Appendix, Fig.~S4 and S5). 
	
	Next, we perform numerical simulations to investigate the nonlinear dynamics of the multistable system over a wider ranger of parameters. Based on the discrete model, we derive the equations of motion (EOMs) of each square in the system. By introducing the following normalized parameters: $K_1=K_s/K_l$, $K_2=K_\theta/(K_la^2)$, $T=t\sqrt{K_l/M}$, $\beta=a\sqrt{M/J}$, $U=u/a$, $V=v/a$ (where $a$ is the distance between the centers of two neighboring squares), we obtain the dimensionless EOMs (SI Appendix, Section~3). We quantify the dynamic response of the system by numerically solving the EOMs, using the fourth order Runge-Kutta method. In the numerical simulations we consider a system of $30\times30$ squares. 
	To trigger a nucleation, we apply rotation $\theta_{in}$ to the $2\times2$ squares at the center, similar to the experiments. We find that, given a proper set of parameters (e.g., in this case $\theta_0=25^{\circ}$, $K_1=0.2$, $K_2=0.0306$, $\beta=3.0556$), there exists a critical angle $\theta_c$. When $\theta_0+\theta_{in}\ge \theta_c$, the nucleation of a new phase occurs, with the $2\times2$ squares transforming from the initial Phase \textit{C} to the new Phase \textit{R}. 
	This transition propagates outward throughout the rest of the metamaterial in the form of a transition wave with some directional dependence 
	(i.e., it travels faster along the $x$ and $y$ axes than along the diagonals; SI Appendix, Fig.~S6). 
	Snapshots from the numerical simulations are displayed in Fig.~\ref{fig:Experimental_results}D for normalized times $T=0$, $40$, $75$, and $100$, showing qualitative agreement with the experimental observations (Movie S2). 
	
	The existence of the critical angle $\theta_c$ suggests that there is an energy threshold $E_c$. To understand the origin of this threshold, we characterize the phase transition observed in our mechanical system from the energy perspective. Analogous to classical first order phase transitions, there is a ``critical nucleus size'' that is required for the new phase to be stable~\cite{onuki2002,Jackson2006}. This results from the competing effects of energy terms that favor the transition (e.g., the energy released by moving from Phase~\textit{C} to Phase~\textit{R} in Fig.~\ref{fig:Experimental_results}B) and terms that do not favor it (e.g., the interface energy between the new phase and the old phase). 
	For the specific system investigated above, we find that the energy threshold is $E_c=1.06\times10^{-2}$ (normalized by $\bar{E}=K_la^2$; see also 
	SI Appendix, Fig.~S7). It is worth noting that the critical nucleus size depends on the choice of parameters. For a different set of parameters, it is possible to obtain a critical nucleus size other than $2\times2$ squares (SI Appendix, Fig.~S8). 
	
	\begin{figure*}[htbp]
		\centerline{ \includegraphics[width=1\textwidth]{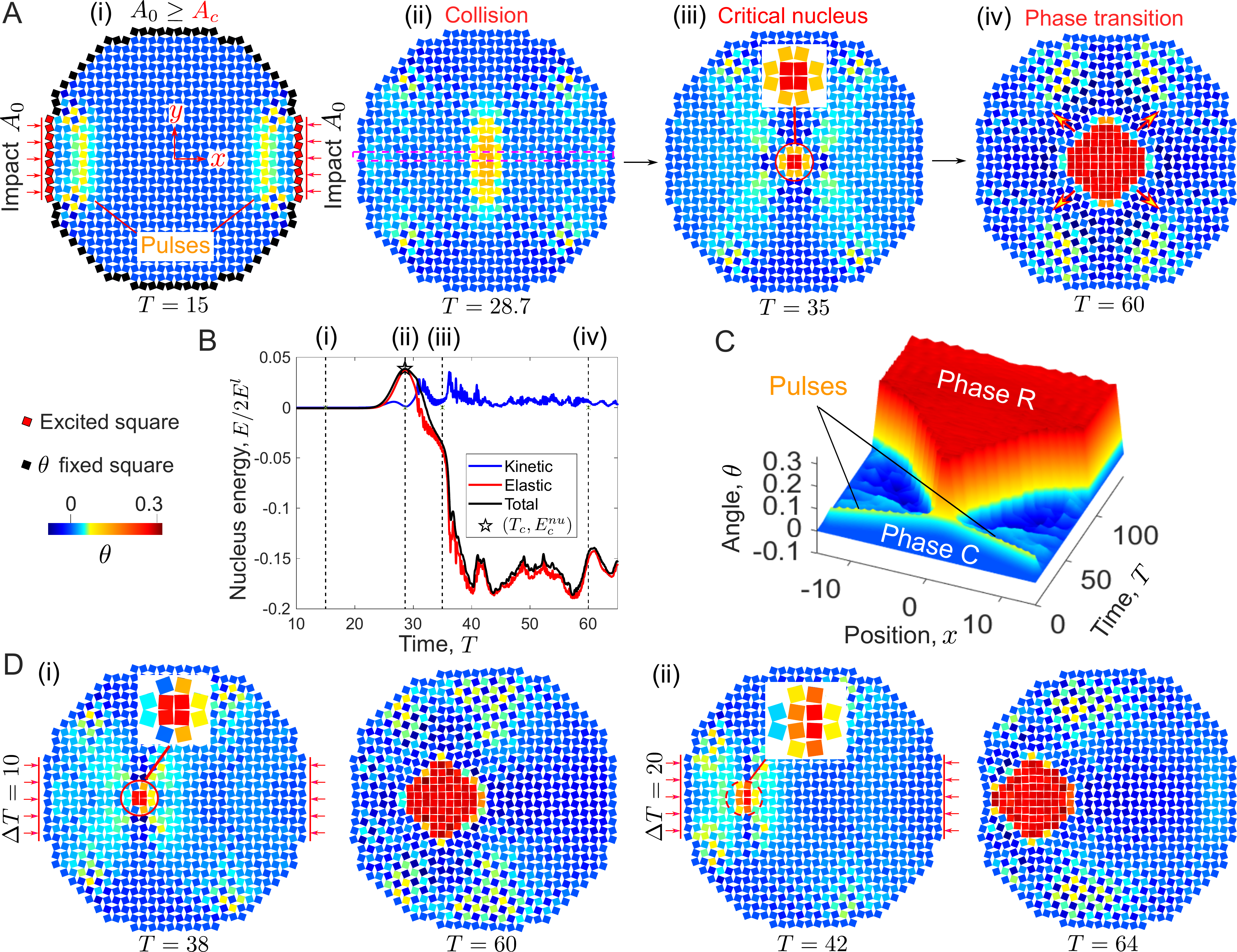}}
		\caption{Head-on collisions of two soliton-like pulses. (A) Snapshots of wavefields for $A_0= 0.306\equiv A_c$: (i) before collision at $T=15$, (ii) during collision at $T=28.7$, (iii) nucleation at $T=35$, and (iv) phase transition at $T=100$. (B) Energy of the cluster at the nucleation site as a function of time, suggesting an energy barrier $E_c^{nu}$ in the total energy curve. (C)~Spatiotemporal plot obtained from the numerical simulation, showing the angle $\theta$ for squares along the propagation direction ($x$ axis) as a function of time. (D)~Control of the location of nucleation via timing of the impulses for (i) $\Delta T=10$ and (ii) $\Delta T=20$, where $\Delta T$ is the time delay of the impact on the left boundary with respect to the impact on the right boundary. 
		}
		\label{fig:Collision_0deg_mode1_v2}
	\end{figure*}
	
	\begin{figure*}[htbp]
		\centerline{ \includegraphics[width=1\textwidth]{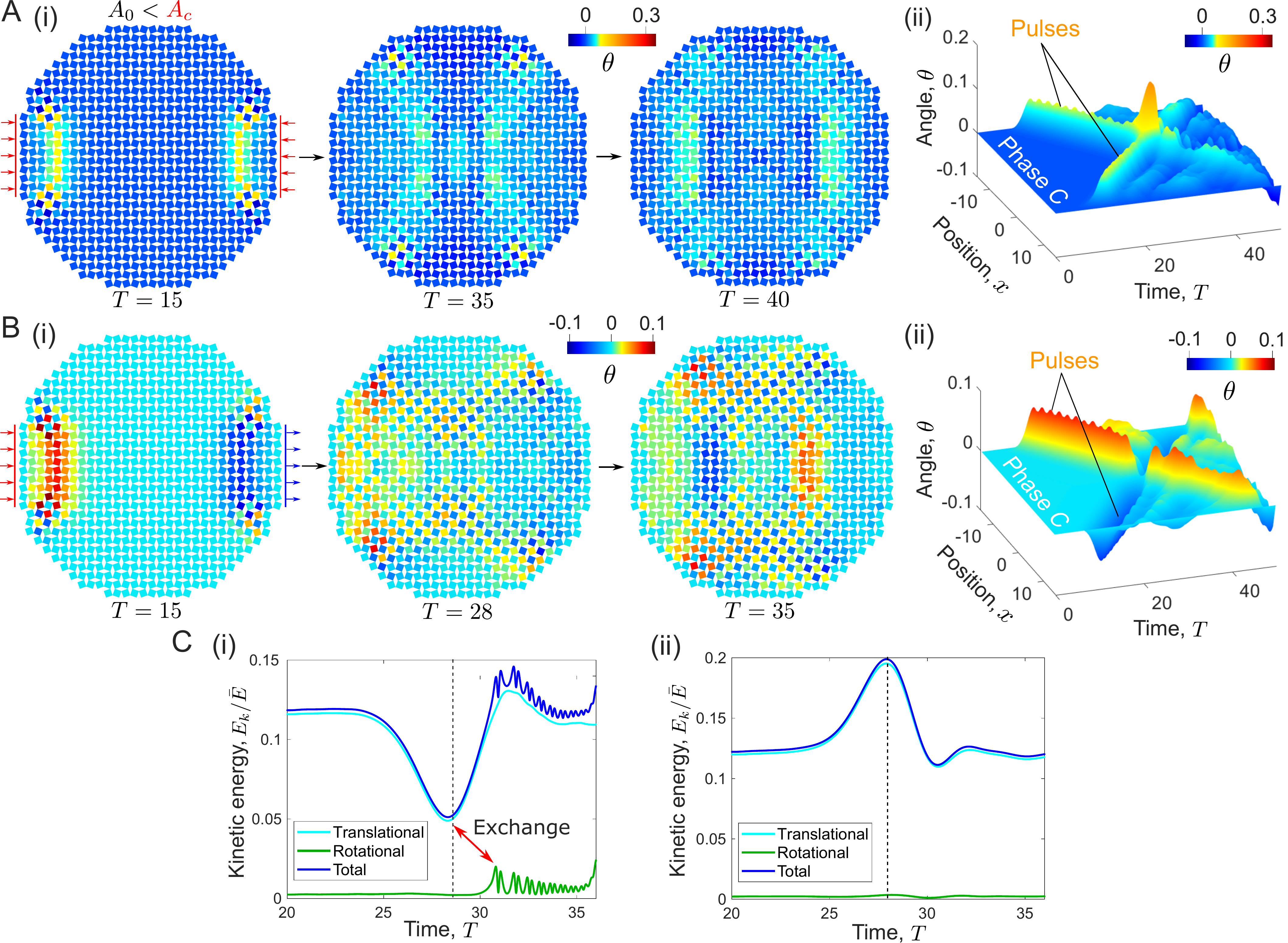}}
		\caption{(A) Head-on collision of two pulses with the same rotational direction for $A_0=0.3<A_c$. (i)~Snapshots of wavefields before collision at $T=15$ and after collision at $T=35$ and $40$, resulting in no phase transition. (ii)~Spatiotemporal plot extracted from the numerical simulation, showing the angle $\theta$ for squares along the propagation direction as a function of time. (B)~Head-on collision of two pulses with the opposite rotational direction for $A_0=A_c$. (i)~Snapshots of wavefields before collision at $T=15$ and after collision at $T=28$ and $35$, showing that the pulses pass through each other. 
			(ii)~Spatiotemporal plot obtained from the numerical simulation, showing the angle $\theta$ for squares along the propagation direction as a function of time. (C)~Kinetic energy of the whole structure as a function of time for a head-on collision of pulses for $A_0=A_c$ with (i)~same rotational directions and (ii)~opposite rotational directions (the vertical dashed lines indicate the time when the two pulses collide); 
			the former case exhibits 
			a significant exchange between the transitional and rotational components of the kinetic energy. } 
		\label{fig:no_nucleation}
	\end{figure*}
	
	\begin{figure*}[htbp]
		\centerline{ \includegraphics[width=1\textwidth]{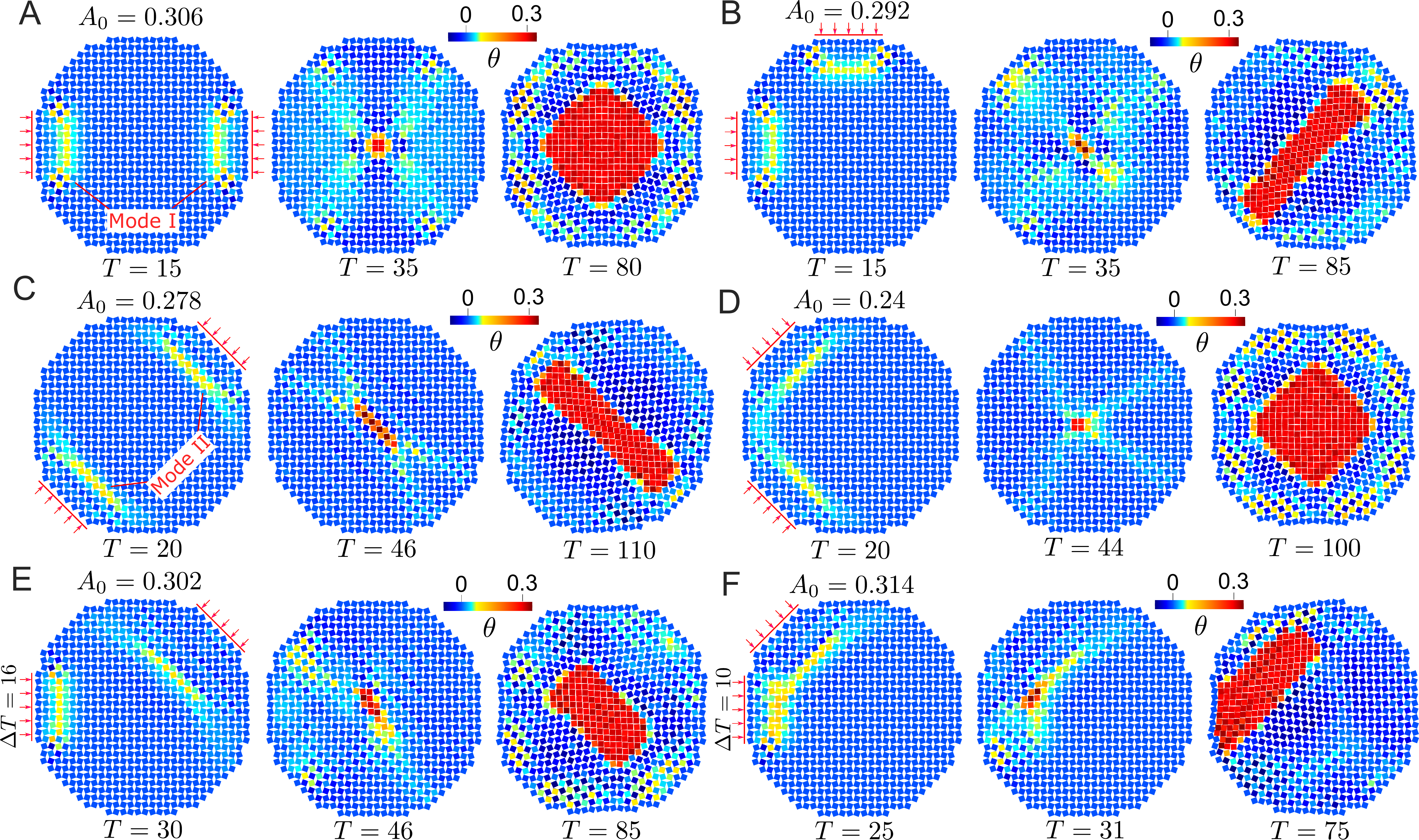}}
		\caption{Different collision scenarios. 
			(A)~Head-on collision of two mode-I pulses with $A_0=0.306$.
			(B)~Perpendicular collision of two mode-I pulses, 
			with $A_0=0.292$. 
			(C)~Head-on collision of two mode-II pulses along the diagonal, with $A_0=0.278$. 
			(D)~Perpendicular collision of two mode-II pulses, 
			with $A_0=0.24$. 
			(E)~Collision of a mode-I pulse and a mode-II pulse propagating along directions oriented $135^{\circ}$ 
			with respect to one another, with $A_0=0.302$. 
			(F)~Collision of a mode-I pulse and a mode-II pulse propagating along directions oriented $45^{\circ}$ 
			with respect to one another, with $A_0=0.314$. 
		} 
		\label{fig:Collision_other}
	\end{figure*}
	
	\section*{Initiating phase transitions via collisions of soliton-like pulses}
	
	Now that we have characterized the energy criteria necessary to induce a phase transition quasistatically, we next consider how a transition could be nucleated by colliding vector solitons. Here, we have intentionally chosen design parameters that produce the smallest critical square nucleus, i.e., $2\times 2$ squares. 
	We consider a circular-shaped system with 30 squares along its diagonal. We impact the sample at different squares along its circumference to initiate pulses that propagate along different directions. Specifically, the impacts are displacement profiles in the form
	\begin{equation}\label{Impact_profile}
		D(T)=\frac{A_0}{2}\tanh \left[(T-T_0)/W\right]+\frac{A_0}{2}\tanh(T_0/W)
	\end{equation}
	where $A_0$ and $W$ are parameters that alter the impact amplitude and shape, respectively. To avoid triggering a nucleation directly at the impacted squares, in the simulations we impose $\theta=0$ to all squares on the boundary. 
	
	\subsection*{Head-on collisions of two pulses with same rotation}
	
	We first investigate head-on collisions of pulses by applying impacts at the left and right boundary. In Fig.~\ref{fig:Collision_0deg_mode1_v2}A, we show snapshots of the wavefields at $T=15$, $28.7$, $35$, and $60$, demonstrating that a phase transition is induced 
	where the two pulses collide (Movie~S3). By sweeping the impact amplitude $A_0$, we identify a critical amplitude $A_c=0.306$, below which a nucleation is not induced by the colliding pulses (see Fig.~\ref{fig:no_nucleation}A and Movie~S4). When $A_0\geq A_c$, the collision of the two pulses can lead to the formation of a critical nucleus. In that case, the new phase propagates outward to the rest of the structure via a transition wave. In Fig.~\ref{fig:Collision_0deg_mode1_v2}B, we plot the normalized energy of the squares at the nucleation site (i.e., the squares 
	in the inset of Fig.~\ref{fig:Collision_0deg_mode1_v2}A(iii)) as a function of time for $A_0=A_c$. We observe that there also exists an energy threshold $E_c^{nu}=3.84\times10^{-2}$ during the collision process. 
	Comparing this energy threshold $E_c^{nu}$ with its counterpart in the previous quasistatic analysis, we note that $E_c^{nu}$ is much larger than $E_c$, a result of the fact that not all of the energy 
	in the propagating pulses will be directed toward forming a new phase during the collision 
	(e.g., some energy is lost in the form of 
	scattered waves). Fig.~\ref{fig:Collision_0deg_mode1_v2}C shows a spatiotemporal plot that provides the angle of the squares 
	along the propagation direction ($x$ axis) as a function of time and position. 
	We also note that the location of nucleation can be changed simply by 
	introducing a time delay $\Delta T$ for the initiation of the impulse on the left with respect to the initiation of the impulse on the right. 
	In Fig.~\ref{fig:Collision_0deg_mode1_v2}D, we demonstrate this by showing snapshots of the simulations for $\Delta T =10$ and $20$ (Movie~S5). 
	
	\subsection*{Head-on collisions of two pulses with opposite rotation}
	
		We also explore head-on collisions of pulses with different rotational directions (Fig.~\ref{fig:no_nucleation}). In contrast with collisions between impulses with the same (positive) rotation (as was triggered by applying two compressive impulses at the left and right boundaries 
		in Fig.~\ref{fig:Collision_0deg_mode1_v2}A) Fig.~\ref{fig:no_nucleation}B shows a collision of two pulses with opposite rotational directions. 
		This is accomplished by changing the excitation at the right boundary from a compressive impact to a tensile impact. 
		The two pulses pass through each other without inducing a nucleation for $A_0=A_c$ (Movie~S6). To better understand this observation, we separate the kinetic energy 
		into two components: one associated with translational motion and the other associated with rotational motion. The results are plotted in Fig.~\ref{fig:no_nucleation}C (i-ii) with $A_0=A_c$ for 
		for same rotation and opposite rotation, respectively. We find that there is some energy exchange between the two kinetic energy components for the same rotation case, i.e., some portion of the translational kinetic energy is transferred to the rotational kinetic energy. However, this energy exchange is almost negligible for the opposite rotation case. This implies that the rotational kinetic energy gained during the collision process is critical for overcoming the energy barrier associated with 
		nucleation. Another interesting scenario is collision of two pulses with negative rotation triggered by two tensile impulses. 
		In this case, the energy exchange is negligible. As a result, nucleation cannot be initiated (SI Appendix, Fig.~S9).
	\subsection*{Effects of propagation distance on nucleation}
	Since the pulses are triggered at the boundary and collide at the center of the structure, it is expected that the propagation distance 
	can affect the wave interactions during the collisions, and therefore may affect the nucleation. We repeat the above analysis for systems with different sizes to examine this effect. The results, as reported in SI Appendix (Fig.~S10), show that the critical amplitude $A_c$ increases significantly as the size increases. We observe dispersion, especially in the direction perpendicular to propagation, which is qualitatively similar to the expected 2D dispersion behavior observed previously \cite{Deng2019}. As a result, its amplitude spatially decays as it propagates through the media. In contrast, the critical energy barrier $E_c^{nu}$ does not change in an appreciable way, which indicates that the energy barrier for inducing a nucleation is a local quantity, and therefore there is no statistically significant change to the energy barrier.  
	
	\subsection*{Collisions of pulses at other angles}
	
	Finally, we consider the effects of propagation direction on the ability of colliding pulses to nucleate a new phase (Fig.~\ref{fig:Collision_other}). 
	The circular shape of the system allows facile excitation of pulses along arbitrary directions of propagation. For example, by applying impacts at the left and top boundary, the two pulses can propagate along both the $x$ and $y$ principal axes (i.e., the positive $x$ direction and the negative $y$ direction, respectively). As shown in Fig.~\ref{fig:Collision_other}B, the two pulses nucleate a new phase during their collision. In this case, the nucleation can be induced at impact amplitude $A_0=0.292$, which is lower than the critical amplitude of a head-on collision (replotted in Fig.~\ref{fig:Collision_other}A). In addition, we observe that, after nucleation, the new phase grows predominantly along the diagonal, at 45$^{\circ}$ 
	relative to the $x$ and $y$ 
	axes. We refer to such pulses, traveling along the $x$ or $y$ axes, as mode~I pulses. Another feasible propagation direction is along the diagonals (referred to as mode~II pulses), a direction previously found to support the propagation of vector solitons in monostable systems of rotating squares~\cite{Deng2019}. 
	Fig.~\ref{fig:Collision_other}C shows a head-on collision between impulses propagating along this direction. 
	Mode-I pulses travel much faster than mode-II pulses under the same impact amplitude, and the wave speeds of both modes slightly decrease as the impact amplitude increases (SI Appendix, Fig.~S12). With the above observations from Fig.~\ref{fig:Collision_other}C, we demonstrate that the head-on collisions of two mode-II pulses can initiate a nucleation with impact amplitude $A_0=0.278$. Then, the new phase grows predominantly along the diagonal at $-45$ degrees. Fig.~\ref{fig:Collision_other}D shows collision of two mode-II pulses propagating along principal axes oriented to one another at $90$ degrees for $A_0=0.24$. Interestingly, we report in Fig.~\ref{fig:Collision_other}E that a mode-I pulse can collide with a mode-II pulse at nearly 135 degrees to initiate a nucleation for $A_0=0.302$ (note that the pulse of mode I is delayed by $\Delta T=16$ to compensate the speed difference between the two modes). Lastly, Fig.~\ref{fig:Collision_other}F shows collision of a mode-I pulse and a mode-II pulse propagating along directions oriented $45$ degrees with respect to one another for $A_0=0.314$ and $\Delta T=10$. It is also worth noting that these various collisions can lead to nucleation with different shape, resulting in rich propagation characteristics of the phase transition as described above.
	
	\section*{Conclusion}
	In conclusion, we have experimentally and numerically investigated phase transitions in macroscopic mechanical metamaterials, analogous to classical solid-solid phase transitions in crystals. First, we have experimentally confirmed and numerically corroborated the existence of phase transitions, which can propagate in the form of transition waves in 2D rotating-squares structures. We have identified the fundamental requirements for inducing nucleation, including the energy threshold and the critical nucleus size. 
	More importantly, 
	we have found a fundamentally new way to initiate these phase transitions, i.e., by colliding two soliton-like pulses. This allows 
	nucleation to occur 
	at arbitrary locations in the metamaterial, which may have significant utility in facile control of shape-morphing structures. 
	Therefore, this work not only contributes fundamentally to the understanding of nonlinear waves, and particularly how collisions of one type of nonlinear wave can induce formation of another type, 
	but could also open new doors for the design of  
	tunable, shape-transforming, and deployable structures. 
	
	\section*{Acknowledgement}
	The authors gratefully acknowledge support via NSF award number 2041410, AFOSR award number FA9550-19-1-0285, DARPA YFA award number W911NF2010278, and the University of Pennsylvania Materials Research Science and
	Engineering Center (MRSEC) (NSF DMR-1720530). H.Y. acknowledges the support of KAKENHI (22K14154).
	
	\bibliographystyle{unsrt}
	\bibliography{pnas_sample}
	
	\clearpage
	\onecolumngrid
	
	\setcounter{figure}{0}
	\setcounter{equation}{0}
	\setcounter{page}{1}
	\renewcommand{\thefigure}{S\arabic{figure}}
	\renewcommand{\theequation}{S\arabic{equation}}
	\renewcommand{\thepage}{S\arabic{page}}
	
	\section*{\Large Supplemental Information}
	\section*{1. Design and Fabrication}
	
\begin{figure}
		\centerline{ \includegraphics[width=\textwidth]{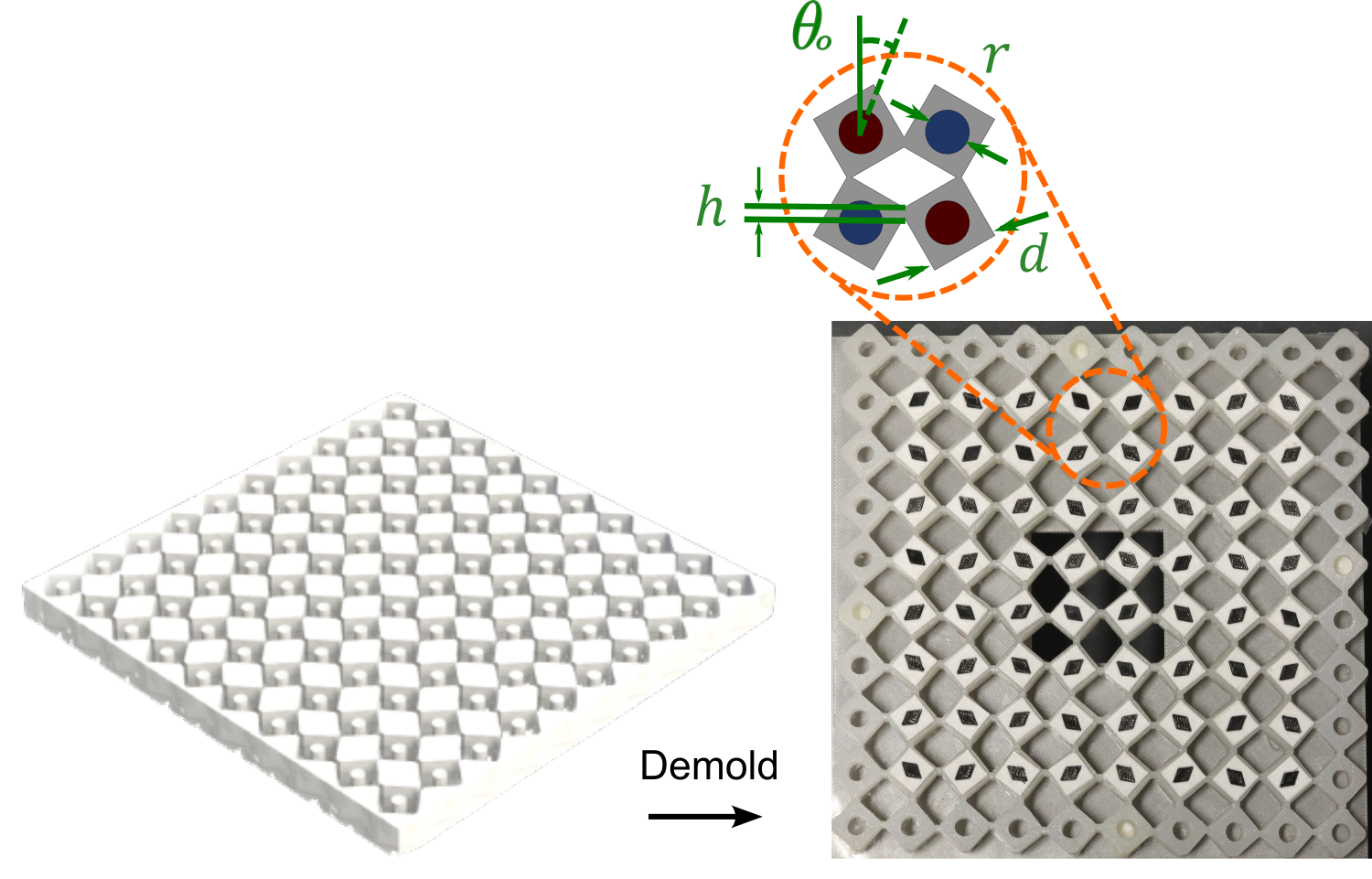}}
		\caption{Fabrication process of a system comprising 10 columns and 10 rows.} 
		\label{fig:SI_fabrication}
	\end{figure}

	In this work, experiments are conducted on building blocks of $2\times2$ elastomeric rotating squares and larger $10\times10$ metamaterials (Fig.~\ref{fig:SI_fabrication}). 
	The squares have edge length $d = 12$~mm and are rotated by an angle $\theta_{Lin} = 5^{\circ}$ with respect to the vertical axis (note that $\theta_{Lin}$ is the initial equilibrium angle without magnets inserted).  We print a mold (MakerGear M2, Polylactic acid (PLA)) with cylindrical extrusions of radius $r = 6$~mm at the center  (Fig.~\ref{fig:SI_fabrication}). Adjacent squares are connected via thin hinges of thickness $h = 1.5$~mm. 
	Silicone (Dragonskin 10, Smooth-On, Inc.) is mixed under vacuum using a Speedmixer (FlackTek, Inc), then poured into the mold and cured at room temperature (six hours). 
	After curing, 
	permanent cylindrical magnets (D41-N52 Neodymium Magnets, K$\&$J Magnetics) are embedded at the center to provide attraction between adjacent squares. Note in Fig.~\ref{fig:SI_fabrication}, magnets are not included in the squares along the edges, to prevent unintended phase changes at the boundary squares that can result from 
	boundary effects. Finally, 3D-printed (MakerGear M2, PLA), diamond-shaped trackers are adhered to the surface of each unit to allow tracking of the nodal rotation during dynamic testing. 
	
	\section*{2. Experiments}
	\subsection*{2.1 Static testing}
\begin{figure}[!htbp]
		\centerline{ \includegraphics[width=1\textwidth]{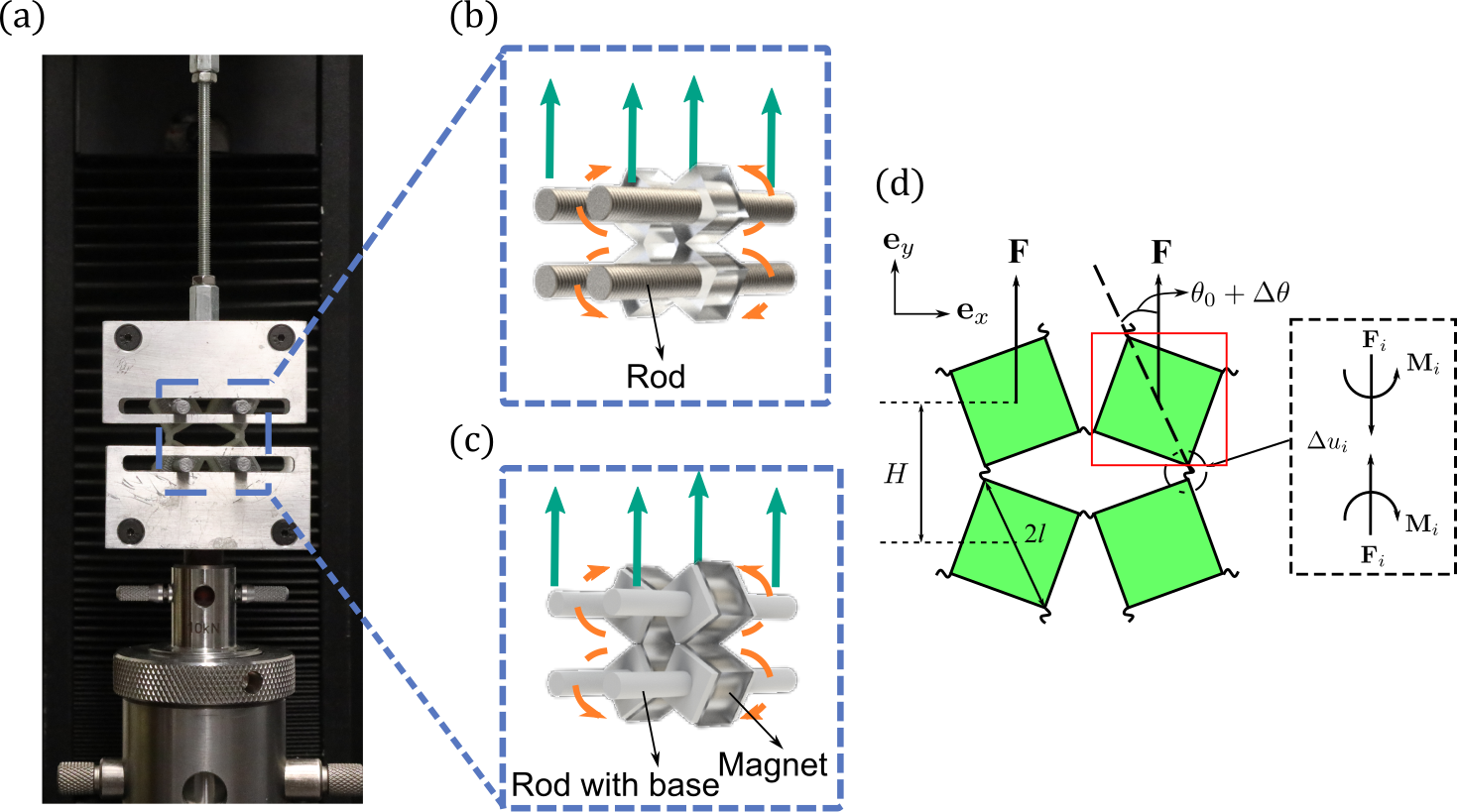}}
		\caption{ (a) Tensile tests were conducted using an Instron model 68SC-5 equipped with a custom aluminum fixture. (b) Schematic of tensile test on a $2\times2$ unit without magnets (the rods are inserted through the holes of the squares). (c) Schematic of tensile test on a $2\times2$ unit with magnets (the rods with bases are attached to both sides of the squares). Green arrows indicate the direction of the applied force; orange arrows indicate the direction of rotation of each square. (d) Schematic of a $2\times2$ unit under tensile loading.} 
		\label{fig:SI_statictest}
	\end{figure}

	To characterize the static properties of the sample, we perform quasistatic tensile tests using an Instron model 68SC-5 in displacement control with a displacement rate of $0.02$~mm/s. Two aluminum fixtures are used to apply displacement to a specimen comprising four squares (two columns), as shown in Fig.~\ref{fig:SI_statictest}(a). 
	Tensile tests are conducted both with and without magnets. 
	
\begin{figure}[htbp]
		\centerline{ \includegraphics[width=1\textwidth]{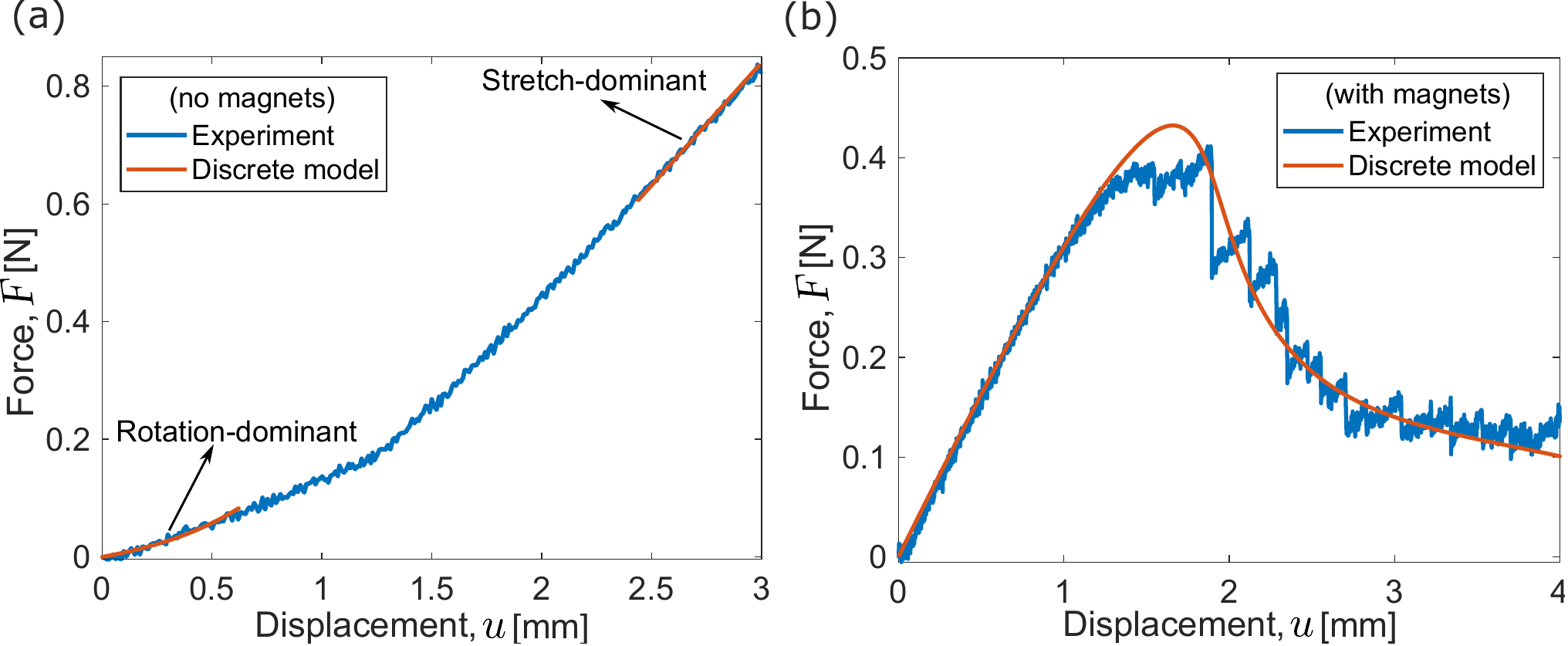}}
		\caption{Force-displacement relationship for the $2\times2$ structures (a) without magnets and (b) with magnets. Blue and red lines indicate results from experiments and discrete model with fitted parameters, respectively.} 
		\label{fig:SI_tensile}
	\end{figure}

	For tests 
	without magnets (Fig.~\ref{fig:SI_statictest}(b)), we embed an aluminum rod at the center of each square. The two ends of the rod maintain alignment via 
	a horizontal slot in the fixture, which allows free rotation and displacement of each square. 
	Fig.~\ref{fig:SI_statictest}(b) and (c) indicate the locations of the applied force (green arrows) and the direction of rotation of each square (orange arrows). 
	Figure~\ref{fig:SI_tensile} shows the measured force-displacement data 
	(blue). 
	
	As discussed in the main text, we introduce a discrete model to capture the behavior of the prototypes. 
	Based on the discrete model (see schematic in Fig.~\ref{fig:SI_statictest}(d)), 
	we can explicitly obtain the force-displacement relationship for a $2\times2$ system under tensile loading (the two squares at the bottom are fixed in the vertical direction but free to rotate). The equations of equilibrium for the square highlighted by the red box can be written as 
	\begin{align}\label{governing_eqns}
		\begin{split}
			\mathbf{F} + \mathbf{F}_i &= \mathbf{0}\\
			2\mathbf{M}_i + \mathbf{r} \times \mathbf{F}_i &=\mathbf{0}
		\end{split}
	\end{align}
	where $\mathbf{r}=\left[l\sin (\theta_0+\Delta\theta) \,\,\, -l\cos (\theta_0+\Delta\theta)\right]^T$, $\mathbf{F}_i=-K_l \Delta u_i \mathbf{e}_y$ and $\mathbf{M}_i=-2K_j \Delta\theta \mathbf{e}_z$ are the longitudinal force and moment of the linkage, respectively. The vertical displacement $u$ (i.e., change in height $H$ defined in Fig.~\ref{fig:SI_statictest}(d)) 
	can be expressed as
	\begin{align}\label{Displacement}
		u=H-H_0=2l\cos(\theta_0+\Delta\theta) -2l\cos \theta_0 + \Delta u_i
	\end{align}
	where $H_0$ is the initial height.
	
	Eq.~\ref{governing_eqns} leads to 
	\begin{align}\label{Force}
		\mathbf{F}=F\mathbf{e}_y=-\frac{4K_j \Delta \theta}{l\sin( \theta_0+\Delta\theta)}\mathbf{e}_y
	\end{align}
	For specimens with magnets, we use the Morse potential to empirically capture the magnetic interactions between squares. In this case, the moment from the hinge $\mathbf{M}_i$ becomes
	\begin{align}\label{Moment_Magnet}
		\mathbf{M}_i=-2K_j \Delta \theta \mathbf{e}_z - T_{Morse}(\Delta{\theta}) 
	\end{align}
	where $T_{Morse}$ is 
	\begin{align}\label{Tmorse}
		\begin{split}
			T_{Morse}(\Delta{\theta}) &= \frac{d V_{Morse}}{d(2\Delta\theta)}\\ &=2\alpha A\Big[e^{4\alpha(\Delta\theta+\theta_{0}-\theta_{Morse})}-e^{2\alpha(\Delta\theta+\theta_{0}-\theta_{Morse})}\Big]\\
			&- 2\alpha A\Big[e^{-4\alpha(\Delta\theta+\theta_{0}+\theta_{Morse})}-e^{-2\alpha(\Delta\theta+\theta_{0}+\theta_{Morse})}\Big]
		\end{split}
	\end{align}
	By fitting the experimental data using Eqs.~\ref{Displacement}-\ref{Moment_Magnet} (red lines in Fig.~\ref{fig:SI_tensile}), we obtain the parameters for the hinge components: $K_j=2.5\times10^{-4}$ for the linear torsional stiffness, and $A=2\times10^{-4}$ and $\alpha=8.5$ for the Morse potential.  With these parameters, we can approximate the multistable energy landscape of the hinge as shown in Fig.~1(b) in the main text. 
	
	\subsection*{2.2 Dynamic testing}
	\label{SI:dynamics}
\begin{figure}[!htbp]
		\centerline{\includegraphics[width=1\textwidth]{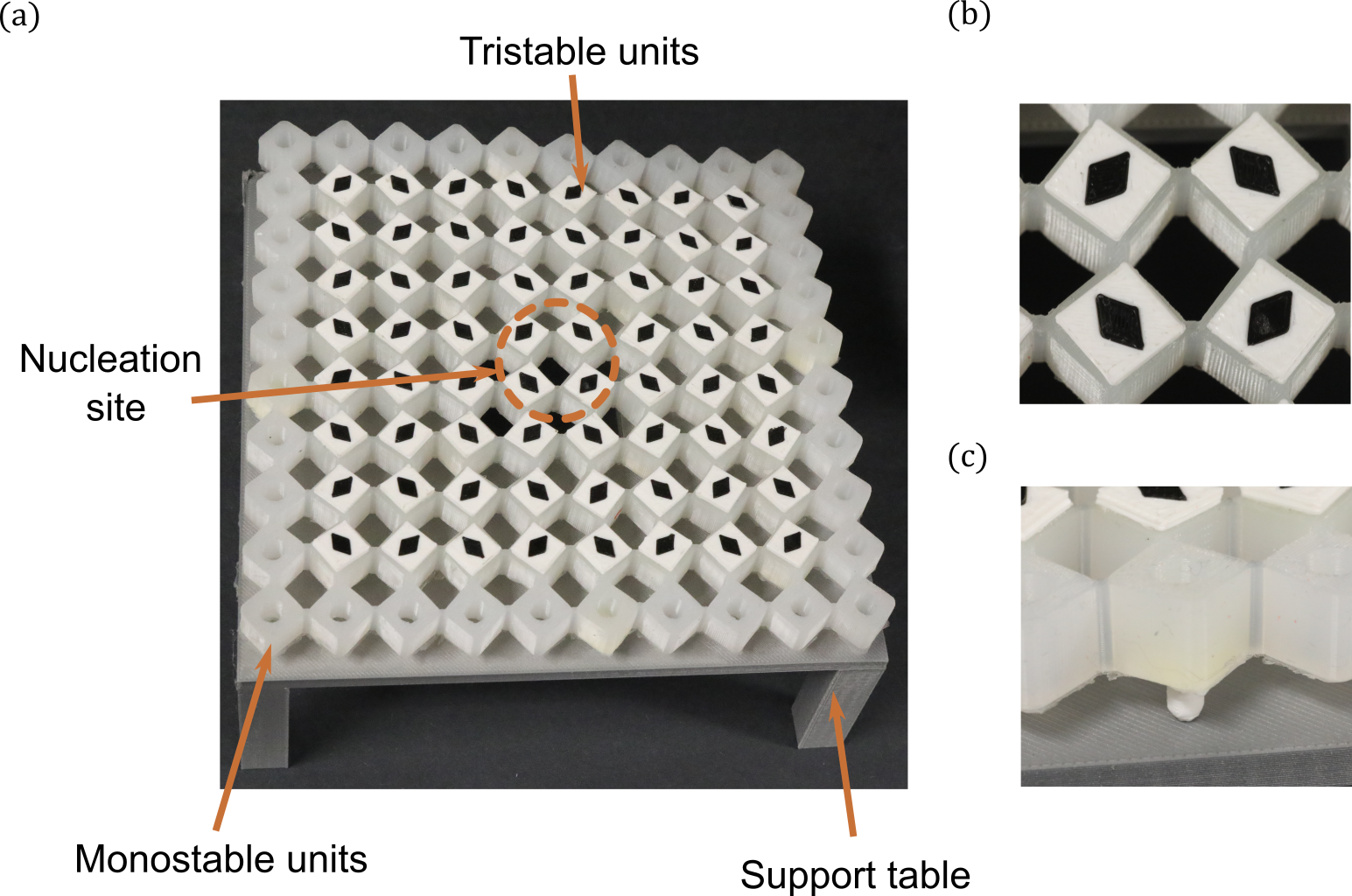}}
		\caption{ (a)~The experimental specimen rests on a support table. The orange circle indicates the nucleation site subjected to loading. (b)~Top view of the center four squares, with diamond-shaped markers adhered to the center of each square. (c)~Detailed view of a square with plastic feet (to reduce friction).}
		\label{fig:SI_dynamictest}
	\end{figure}

\begin{figure}[htbp]
		\centerline{ \includegraphics[width=0.8\textwidth]{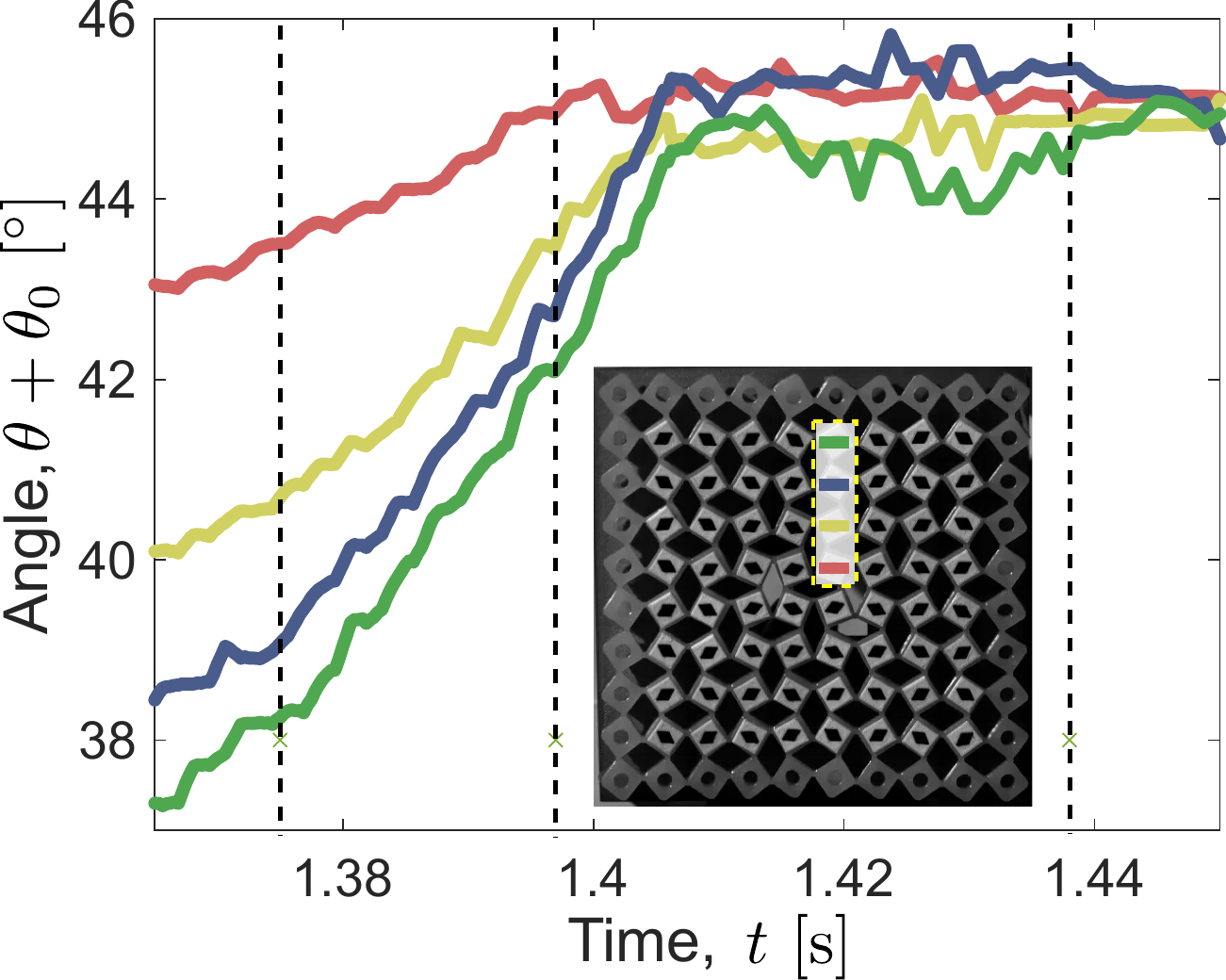}}
		\caption{Experimentally measured angles of the four squares indicated by the yellow dashed lines in the inset. The three vertical lines correspond to the times of the later three optical images displayed in Fig.1(c) in the main text.} 
		\label{fig:Dynamic_test_results}
	\end{figure}

	To experimentally demonstrate phase transformations, 
	we use a 10-column by 10-row sample on a plastic surface (Fig.~\ref{fig:SI_dynamictest}(a)). Quasistatic loading is applied to the two vertical hinges connecting the center four squares at the nucleation site. Note, the squares at the edges do not have magnets, to prevent unintended nucleation at the edges induced by boundary effects. Figure~\ref{fig:SI_dynamictest}(b) and (c) show a detailed view of the center four squares and friction-reducing feet 
	(MakerGear M2, PLA), respectively. The phase transformation is recorded using 
	a high-speed camera (Photron FASTCAM Mini AX) 
	at 6400~frames per second. Diamond-shaped markers are placed at 
	the center of each square to allow tracking of the rotation and displacement of the squares, using a custom Python script (Fig.~\ref{fig:SI_dynamictest}(b)). In Fig.~\ref{fig:Dynamic_test_results}, we plot the experimentally measured angles of the four squares highlighted in the inset of Fig.~\ref{fig:Dynamic_test_results}, showing the transition from the initial phase to the new phase (i.e., Phase~\textit{R}, with $\theta_R \approx45^{\circ}$).
	
	\section*{3. Equations of Motion}
	
	Based on the discrete model introduced in the main text, the Hamiltonian of a 2D rotating-squares system can be written as
	\begin{align}\label{Hamiltonian}
		\begin{split}
			H&=\frac{1}{2} \sum_{n,m} \left( M \dot u_{n,m}^2+M \dot v_{n,m}^2+J\dot \theta_{n,m}^2\right) +\frac{1}{2} K_\theta\sum_{n,m} \Big[\left(\theta_{n,m} + \theta_{n-1,m}\right)^2+(\theta_{n,m}+ \theta_{n,m-1})^2\Big] \\
			&+\frac{1}{2} K_l\sum_{n,m}\Big[ u_{n,m}-u_{n-1,m} +L\cos(\theta_{n,m}+\theta_0)-L\cos(\theta_{n-1,m}+\theta_0)\Big]^2\\
			&+\frac{1}{2} K_l\sum_{n,m}\Big[ v_{n,m}-v_{n,m-1} +L\cos(\theta_{n,m}+\theta_0)-L\cos(\theta_{n,m-1}+\theta_0)\Big]^2\\
			&+\frac{1}{2} K_s\sum_{n,m}\Big[ u_{n,m}-u_{n,m-1} +(-1)^{n+m}L\sin(\theta_{n,m}+\theta_0)-(-1)^{n+m-1}L\sin(\theta_{n,m-1}+\theta_0)\Big]^2\\
			&+\frac{1}{2} K_s\sum_{n,m}\Big[ v_{n,m}-v_{n-1,m} -(-1)^{n+m}L\sin(\theta_{n,m}+\theta_0)+(-1)^{n+m-1}L\sin(\theta_{n-1,m}+\theta_0)\Big]^2,
		\end{split}
	\end{align}
	where $L=\frac{a}{2\cos \theta_0}$ is half of the diagonal length of the square. Then, Hamilton's equations read
	\begin{align}
		&\begin{aligned}
			M\ddot{u}_{n,m}&=-\frac{\partial H}{\partial u_{n,m}}
		\end{aligned},\\
		&\begin{aligned}
			M\ddot{v}_{n,m}&=-\frac{\partial H}{\partial v_{n,m}}
		\end{aligned},\\
		&\begin{aligned}\label{Hamilton_eq3}
			J\ddot{\theta}_{n,m}&=-\frac{\partial H}{\partial \theta_{n,m}}.
		\end{aligned}
	\end{align}
	
	From Eq.~\ref{Hamiltonian} to Eq.~\ref{Hamilton_eq3}, the equations of motion (EOMs) for the square at site $(n,m)$ 
	can be derived as
	\begin{align}\label{governing_eqns_u}
		\begin{split}
			M \frac{\partial^2 u_{n,m}}{\partial t^2} 
			&= K_l \left(u_{n-1,m} +u_{n+1,m}-2u_{n,m}) + K_s(u_{n,m-1} +u_{n,m+1}-2u_{n,m}\right)\\&+K_l\frac{a}{2\cos \theta_0}\Big[\cos(\theta_{n-1,m}+\theta_0)-\cos(\theta_{n+1,m}+\theta_0)\Big]\\
			&+(-1)^{n+m}K_s\frac{a}{2\cos \theta_0}\Big[\sin(\theta_{n,m+1}+\theta_0)- \sin(\theta_{n,m-1}+\theta_0)\Big],
		\end{split}
	\end{align}
	
	\begin{align}\label{governing_eqns_v}
		\begin{split}
			M \frac{\partial^2 v_{n,m}}{\partial t^2}
			&= K_s (v_{n-1,m} +v_{n+1,m}-2v_{n,m}) + K_l(v_{n,m-1} +v_{n,m+1}-2v_{n,m})\\
			&+K_l\frac{a}{2\cos \theta_0}\Big[\cos(\theta_{n,m-1}+\theta_0)-\cos(\theta_{n,m+1}+\theta_0)\Big]\\
			&+(-1)^{n+m}K_s\frac{a}{2\cos \theta_0}\Big[-\sin(\theta_{n+1,m}
			+\theta_0)+\sin(\theta_{n-1,m}+\theta_0)\Big],
		\end{split}
	\end{align}
	
	\begin{align}\label{governing_eqns_theta}
		\begin{split}
			J \frac{\partial^2 \theta_{n,m}}{\partial t^2} &= -K_\theta (\theta_{n-1,m} +\theta_{n+1,m}+\theta_{n,m+1} +\theta_{n,m-1}+4\theta_{n,m}+8\theta_0-8\theta_{Lin})\\
			&-K_l\frac{a}{2\cos \theta_0}\sin(\theta_{n,m}+\theta_0)\left(u_{n+1,m}+v_{n,m+1}-u_{n-1,m}-v_{n,m-1}\right)\\
			&-K_l\frac{a^2}{4\cos^2 \theta_0}\sin(\theta_{n,m}+\theta_0)\Big[\cos(\theta_{n+1,m}+\theta_0)-\cos(\theta_{n-1,m}+\theta_0)\\
			&-\cos(\theta_{n,m+1}+\theta_0)-\cos(\theta_{n,m-1}+\theta_0)-4\cos(\theta_{n,m}+\theta_0)+8\cos\theta_0\Big]\\
			&+(-1)^{n+m}K_s\frac{a}{2\cos \theta_0}\cos(\theta_{n,m}+\theta_0)
			\left(u_{n,m+1}-u_{n,m-1}+v_{n-1,m}-v_{n+1,m}\right)\\
			&+K_s\frac{a^2}{4\cos^2 \theta_0}\cos(\theta_{n,m}+\theta_0)\Big[\sin(\theta_{n+1,m}+\theta_0)+\sin(\theta_{n,m+1}+\theta_0)\\
			&-4\sin(\theta_{n,m}+\theta_0)+\sin(\theta_{n-1,m}+\theta_0)+\sin(\theta_{n,m-1}+\theta_0)\Big]\\
			&-T_{Morse}(\Delta{\theta_{n+1,m}})-T_{Morse}(\Delta{\theta_{n-1,m}})-T_{Morse}(\Delta{\theta_{n,m+1}})-T_{Morse}(\Delta{\theta_{n,m-1}}),
		\end{split}
	\end{align}
	where $\Delta{\theta_{n\pm1,m\pm1}}=\theta_{n,m}+\theta_{n\pm1,m\pm1}+2(\theta_0-\theta_{Lin})$. Note, 
	we define the positive direction of rotation with alternating sign 
	for neighboring squares.
	
	By introducing $K_1=K_s/K_l$, $K_2=k_\theta/(K_la^2)$, $T=t\sqrt{K_l/M}$, $\beta=a\sqrt{M/J}$, $U=u/a$, $V=v/a$, we can obtain the following dimensionless EOMs
	\begin{align}\label{governing_eqns_U}
		\begin{split}
			\frac{\partial^2 U_{n,m}}{\partial T^2}
			&=  (U_{n-1,m} +U_{n+1,m}-2U_{n,m}) + K_1(U_{n,m-1} +U_{n,m+1}-2U_{n,m})\\&+\frac{1}{2\cos \theta_0}\Big[\cos(\theta_{n-1,m}+\theta_0)-\cos(\theta_{n+1,m}+\theta_0)\Big]\\&+(-1)^{n+m}\frac{K_1}{2\cos \theta_0}\Big[\sin(\theta_{n,m+1}+\theta_0)- \sin(\theta_{n,m-1}+\theta_0)\Big],
		\end{split}
	\end{align}
	\begin{align}\label{governing_eqns_V}
		\begin{split}
			\frac{\partial^2 V_{n,m}}{\partial T^2}&= K_1 (V_{n-1,m} +V_{n+1,m}-2V_{n,m}) + (V_{n,m-1} +V_{n,m+1}-2V_{n,m})\\
			&+\frac{1}{2\cos \theta_0}\Big[\cos(\theta_{n,m-1}+\theta_0)-\cos(\theta_{n,m+1}+\theta_0)\Big]\\&+(-1)^{n+m}\frac{K_1}{2\cos \theta_0}\Big[-\sin(\theta_{n+1,m}+\theta_0)+\sin(\theta_{n-1,m}+\theta_0)\Big],
		\end{split}
	\end{align}
	
	\begin{align}\label{governing_eqns_Theta}
		\begin{split}
			\frac{1}{\beta^2} \frac{\partial^2 \theta_{n,m}}{\partial T^2} &= -K_2 (\theta_{n-1,m} +\theta_{n+1,m}+\theta_{n,m+1} +\theta_{n,m-1}+4\theta_{n,m}+8\theta_0-8\theta_{Lin})\\
			&-\frac{\sin(\theta_{n,m}+\theta_0)}{4\cos^2 \theta_0}[2\cos \theta_0(U_{n+1,m}+V_{n,m+1}-U_{n-1,m}-V_{n,m-1})\\&-\cos(\theta_{n+1,m}+\theta_0)-\cos(\theta_{n-1,m}+\theta_0)-\cos(\theta_{n,m+1}+\theta_0)\\&-\cos(\theta_{n,m-1}+\theta_0)-4\cos(\theta_{n,m}+\theta_0)+8\cos\theta_0]\\
			&+\frac{(-1)^{n+m}K_1}{4\cos^2 \theta_0}\cos(\theta_{n,m}+\theta_0)\Big[2\cos \theta_0(U_{n,m+1}-U_{n,m-1}+V_{n-1,m}-V_{n+1,m})\Big]\\
			&+\frac{K_1}{4\cos^2 \theta_0}\cos(\theta_{n,m}+\theta_0)\Big[\sin(\theta_{n+1,m}+\theta_0)+\sin(\theta_{n,m+1}+\theta_0)\\
			&-4\sin(\theta_{n,m}+\theta_0)+\sin(\theta_{n-1,m}+\theta_0)+\sin(\theta_{n,m-1}+\theta_0)\Big]\\
			&-\bar{T}_{Morse}(\Delta{\theta_{n+1,m}})-\bar{T}_{Morse}(\Delta{\theta_{n-1,m}})-\bar{T}_{Morse}(\Delta{\theta_{n,m+1}})-\bar{T}_{Morse}(\Delta{\theta_{n,m-1}}),
		\end{split}
	\end{align}
	where $\bar{T}_{Morse}=T_{Morse}/(K_la^2)$.
	The dimensionless EOMs of the system can be obtained by considering Eqs.~\ref{governing_eqns_U}-\ref{governing_eqns_Theta} for all squares. Then, full-scale simulations can be conducted by numerically solving the system's EOMs using the fourth order Runge-Kutta method (via the Matlab function \textit{ode45}). Based on preliminary numerical results, we observe that, after a transition wave is initiated, squares in the new phase can undergo large oscillations due to the energy release from the initial Phase~$C$ to the new  Phase~\textit{R}. To account for the disspision observed in the experiments, we include damping in the simulations by introducing the following simple viscous damping terms in the EOMs: $c_u=\lambda_u \frac{\partial U}{\partial T}$, $c_v=\lambda_v \frac{\partial V}{\partial T}$, and $c_\theta=\lambda_\theta \frac{\partial \theta}{\partial T}$, in which $\lambda_u$ and $\lambda_v$ are damping coefficients for translational motion in the $x$ and $y$ directions, respectively, and $\lambda_\theta$ is the damping coefficient for rotational motion. 
	Damping is added only after the new phase is formed in the simulations (put numbers).
	
	\section*{4. Numerical Characterization}
	\subsection*{4.1 Anisotropy of the 2D transition wave}
\begin{figure*}[htbp]
		\centerline{ \includegraphics[width=1\textwidth]{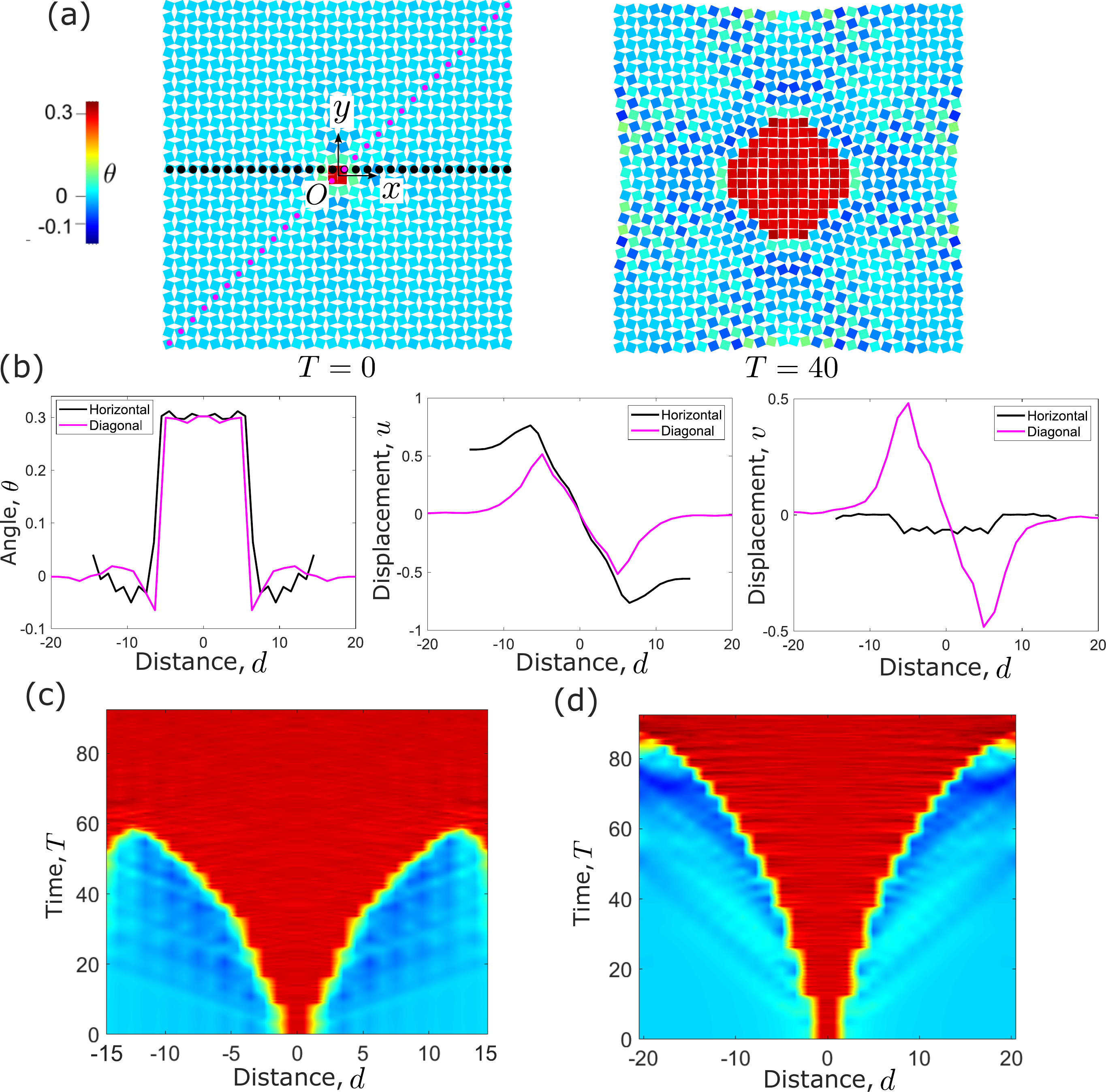}}
		\caption{Anisotropy of the transition wave. (a) Snapshots of the wavefield at $T=0$ ($T$ is set to 0 when the $2\times2$ squares at the center are rotated to Phase~\textit{R}) and $40$. (b)~Spatial profiles along the horizontal (black dots) and diagonal (magenta dots) directions for angle $\theta$, displacement $u$, and displacement $v$. Contour plots of the spatio-temporal data for the angle along the (c) horizontal and (d) diagonal directions.   } 
		\label{fig:Transition_wave_mode_analysis}
	\end{figure*}

	As reported in the main text (Fig.~1(d)), a transition wave triggered at the center of our system propagates outward anisotropically. The wave fronts propagate along the diagonals of the system (i.e., $\pm 45^{\circ}$ with respect to the $x$ axis). To further corroborate this observation, we extract and plot in Fig.~\ref{fig:Transition_wave_mode_analysis}(b) the spatial profiles at $T=40$ for all three degrees of freedom (i.e., displacement $u$ and $v$, and 
	angle $\theta$) along the horizontal and diagonal directions, as indicated by the black and magenta dots in Fig.~\ref{fig:Transition_wave_mode_analysis}(a), respectively. In Fig.~\ref{fig:Transition_wave_mode_analysis}(c) and (d), we display the contour plots of the spatio-temporal data of the angles along the horizontal and diagonal directions, respectively. The transition wave propagates considerably faster along the horizontal direction.
	
	\subsection*{4.2 Energy threshold for inducing a nucleation quasistatically}
\begin{figure*}[htbp]
		\centerline{ \includegraphics[width=1\textwidth]{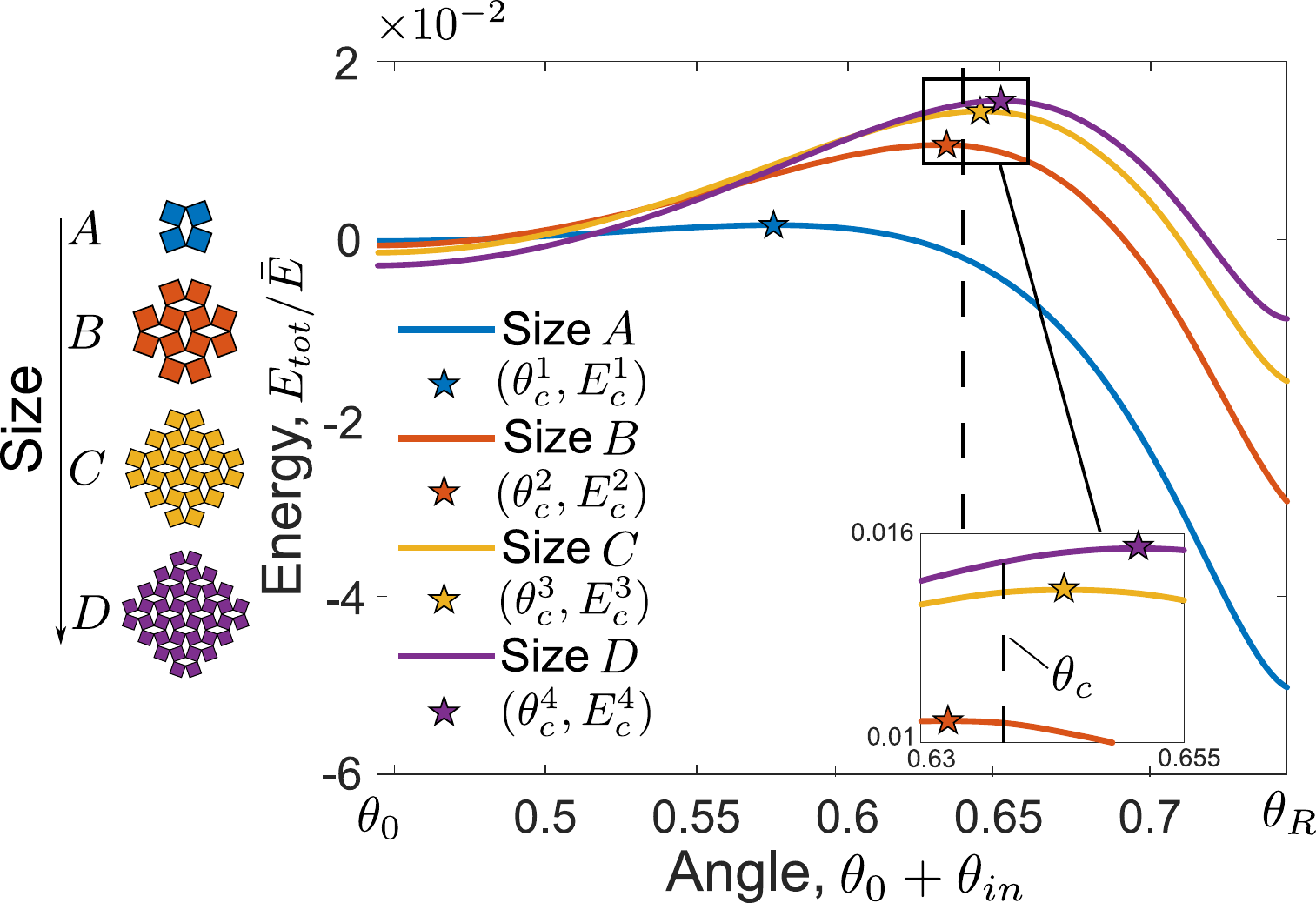}}
		\caption{ Normalized total energy as a function of angle $\theta_0+\theta_{in}$ during the whole quasistatic loading process (i.e., until $\theta_0+\theta_{in}=\theta_R$) for four different, but concentric, square clusters \textit{A} (this is where the rotation $\theta_{in}$ is applied), \textit{B}, \textit{C}, and \textit{D}. The energy curves feature distinct energy barriers indicated by the stars. The critical loading angle is indicated by the vertical dashed line, suggesting a nucleation can be induced after cluster \textit{B} overcomes its energy barrier $E_c^2$. } 
		\label{fig:Quasistatic_analysis}
	\end{figure*}

	As discussed in the main text, the existence of the critical angle $\theta_c$ suggests that there is an energy threshold $E_c$. Once the energy threshold is reached, 
	a 
	critical nucleus can be formed (i.e., $2\times2$ squares in the new phase~\textit{R}). 
	We consider four different, but concentric, square clusters \textit{A} (this is where the rotations are applied), \textit{B}, \textit{C}, and \textit{D}, as shown in Fig.~\ref{fig:Quasistatic_analysis}. Then, we plot the dimensionless energy (normalized by $\bar{E}=K_la^2$) as a function of angle $\theta_0+\theta_{in}$ for the four clusters during the whole quasistatic loading process (i.e., until cluster \textit{A} is fully rotated into the new Phase~\textit{R}). we note that, when the quasistatic loading is present, a phase transition cannot be induced before cluster \textit{A} is fully in Phase~\textit{R}. We indicate the critical angle $\theta_c$ by the vertical dashed line. Clearly, each cluster features an energy barrier $E_c^i$ at a certain angle $\theta_c^i$, in which $i=1,\,2,\,3,\,4$ correspond to cluster \textit{A}, \textit{B}, \textit{C}, and \textit{D}, respectively. Moreover, we note that the critical angle $\theta_c$ is located between $\theta_c^2$ and $\theta_c^3$, which implies that a nucleation can be triggered after cluster \textit{B} overcomes its energy barrier $E_c^2$. Thus, the energy threshold for inducing a nucleation under quasistatic loading conditions is identified as $E_c=E_c^2$.
	
	\subsection*{4.3 Numerical determination of critical nucleus size}
\begin{figure*}[htbp]
		\centerline{ \includegraphics[width=1\textwidth]{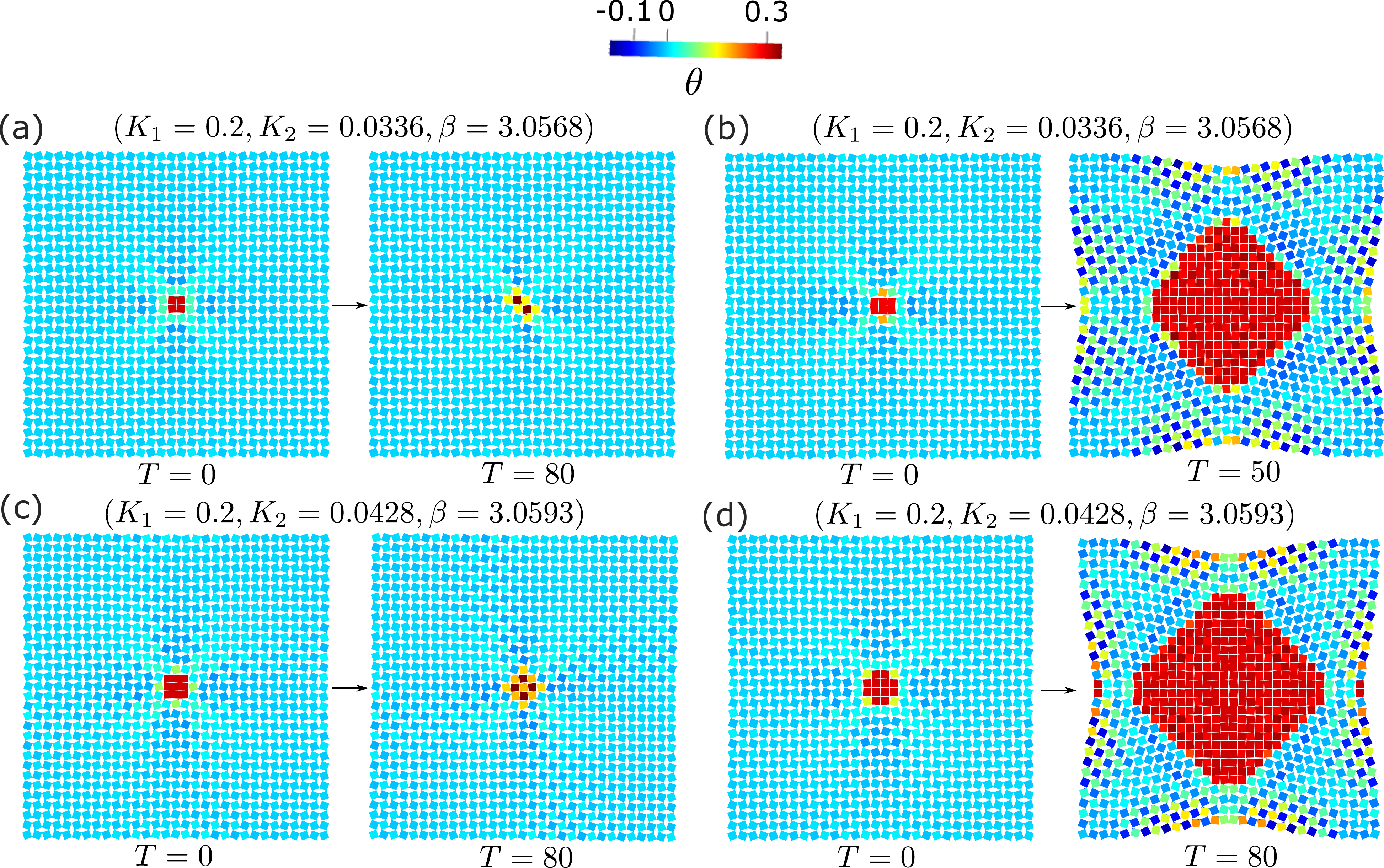}}
		\caption{Critical nucleus size for two sets of parameters: (a) and (b) 6 squares with a rectangular shape for  $(K_1=0.2,K_2=0.0336,\beta=3.0568)$; (c) and (d) $12$ squares with a ``+" shape for $(K_1=0.2,K_2=0.0428,\beta=3.0593)$.   } 
		\label{fig:Nucleus_size_SI}
	\end{figure*}

	Fig.~\ref{fig:Nucleus_size_SI} shows how we determine the critical nucleus size via full-scale simulations for two other sets of parameters. Specifically, we sweep the size of squares that are quasistatically rotated into the new phase (Phase~\textit{R}), starting from $2\times2$ squares at the center, until a phase transition is triggered and then propagates. For ($K_1=0.2$, $K_2=0.0336$, $\beta=3.0568$), we numerically determine the critical nucleus size as 6 squares with a rectangular shape as shown in Fig.~\ref{fig:Nucleus_size_SI}(b). For ($K_1=0.2$, $K_2=0.0428$, $\beta=3.0593$), we numerically determine the critical nucleus size as 12 squares with a ``+" shape as shown in Fig.~\ref{fig:Nucleus_size_SI}(d). We note that this numerical approach becomes inefficient in cases where a set of parameters leads to a large critical nucleus size. 
	
	\subsection*{4.4 Head-on collision of two pulses triggered by tensile impulses}
	
\begin{figure*}[htbp]
		\centerline{ \includegraphics[width=1\textwidth]{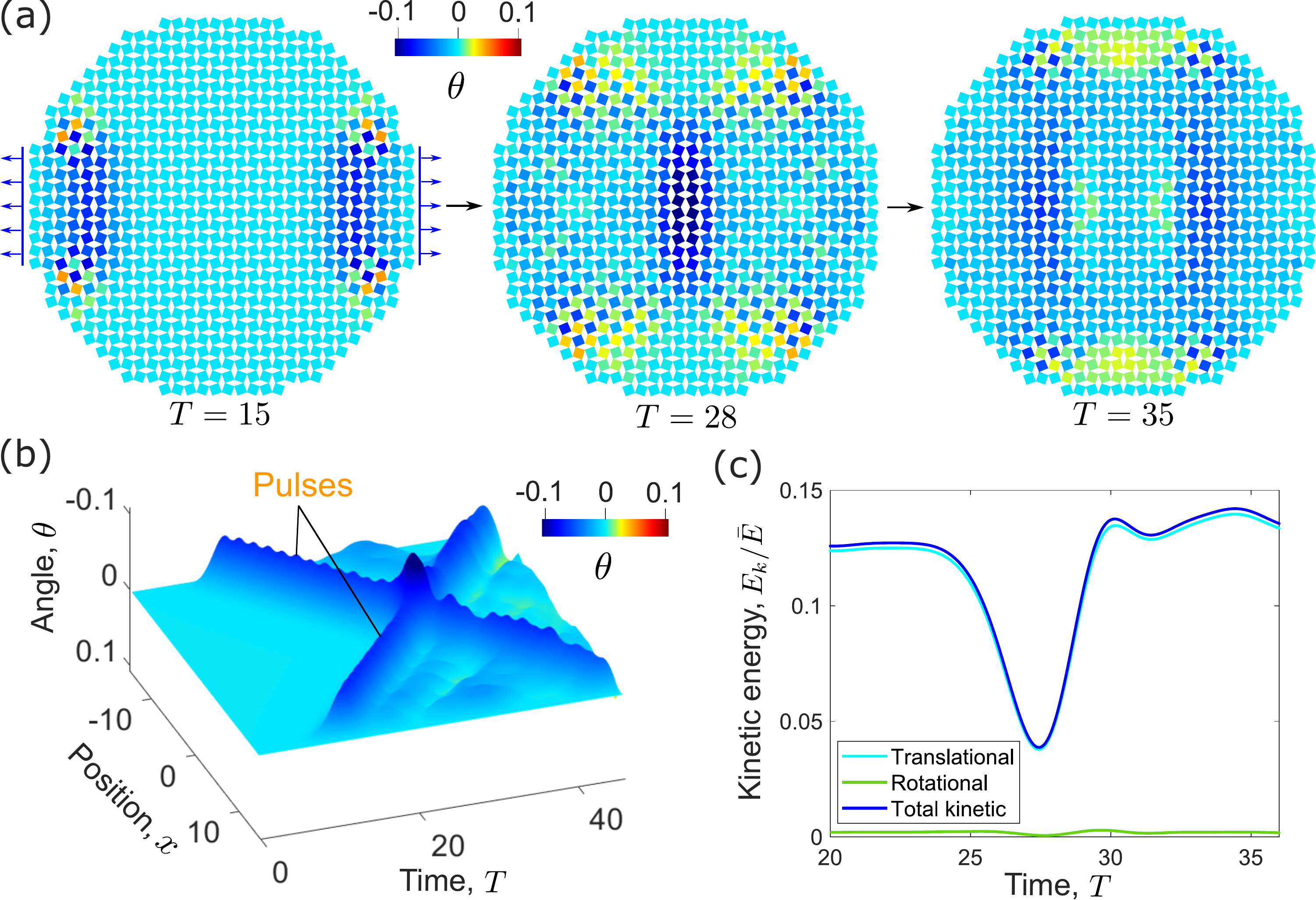}}
		\caption{ Head-on collision of two pulses triggered by tensile impulses. 
			(a) Snapshots of the wavefield before collision at $T=15$, during collision at $T=28$, and after collision at $T=35$. (b) Spatiotemporal plot obtained from the numerical simulation, showing the angle $\theta$ for squares along the propagation direction as a function of time. (c) Kinetic energy of the whole structure as a function of time. } 
		\label{fig:Collision_0deg_mode1_pulling}
	\end{figure*}

	We show in Fig.~\ref{fig:Collision_0deg_mode1_pulling} the simulation result for a head-on collision of two pulses with same (negative) rotation, which is obtained with two tensile impulses 
	for $A_0=0.306$. Fig.~\ref{fig:Collision_0deg_mode1_pulling}(a) displays snapshots of the wavefield before collision at $T=15$, during collision at $T=28$, and after collision at $T=35$. Fig.~\ref{fig:Nucleus_size_SI}(b) gives a spatiotemporal plot of the angle of the squares extracted along the propagation direction, and Fig.~\ref{fig:Collision_0deg_mode1_pulling}(c) gives the total kinetic energy of the system as a function of time. In this case, nucleation does not occur and the energy exchange between the the two components(i.e., translational and rotational) of the kinetic energy is negligible.
	
	\subsection*{4.5 Effect of propagation distance on collision-induced nucleation}
\begin{figure*}[htbp]
		\centerline{ \includegraphics[width=1\textwidth]{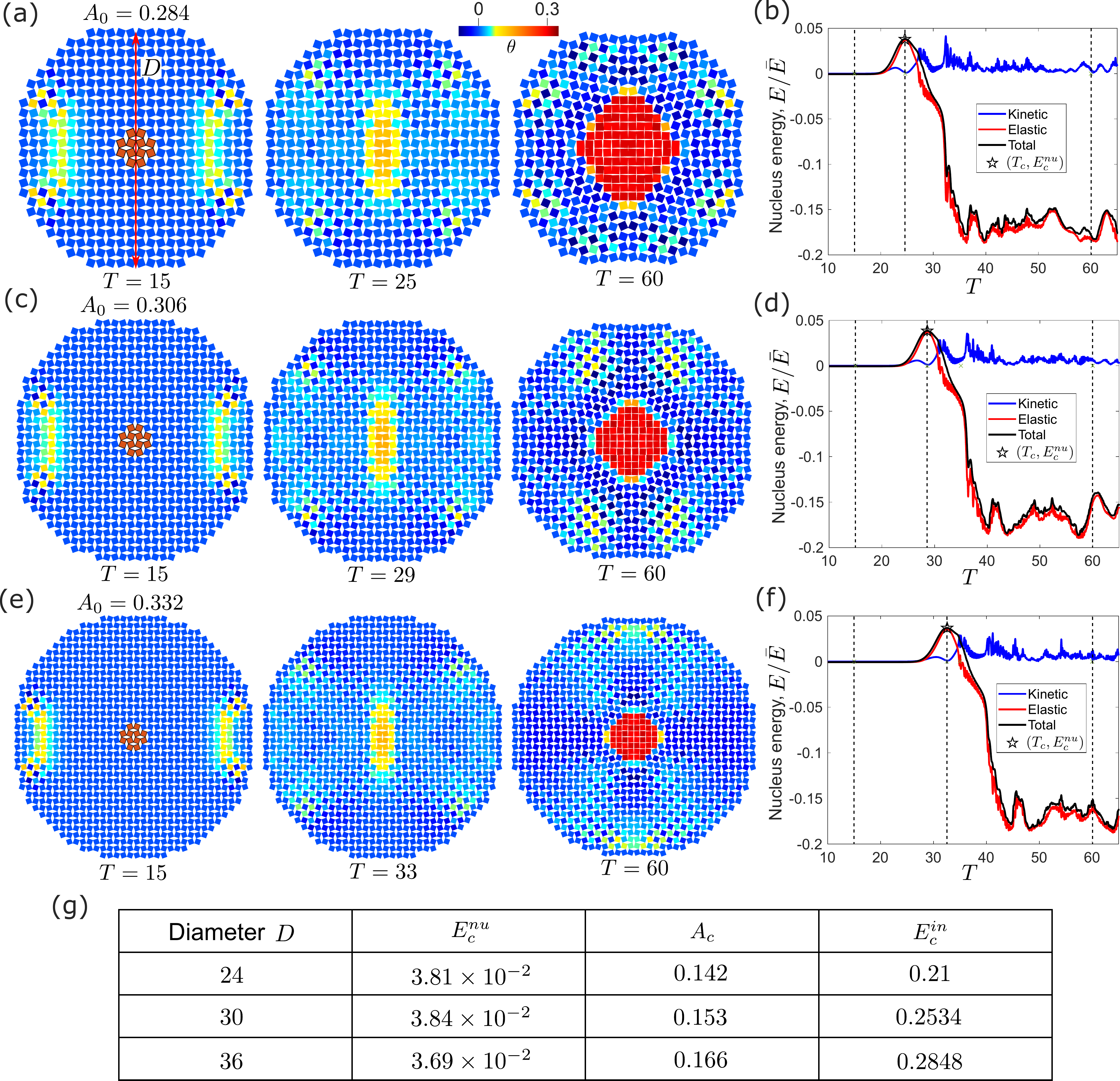}}
		\caption{Head-on collisions of nonlinear pulses in circular systems for three different diameters. Snapshots of wavefields and critical energy barrier of the nucleus highlighted in maroon for: (a) and (b) $D=24$ and $A_0=0.142$; (c) and (d) $D=30$ and $A_0=0.153$; (e) and (f) $D=24$ and $A_0=0.142$. (g) Table: critical energy barrier $E_c^{nu}$, critical impact amplitude $A_c$, and critical total input energy $E_c^{in}$ for $D=24$, $30$, and $36$.} 
		\label{fig:Collision_0deg_mode1}
	\end{figure*}

	To explore the effect of propagation distance on collision-induced nucleation, 
	we consider three circular systems with different diameters $D=24$, $30$, and $36$ (note that $D=30$ is the reference case studied in the main text). In Fig.~\ref{fig:Collision_0deg_mode1}(a)-(f), we report the snapshots of the wavefields and the energy of the nucleus highlighted in maroon for $D=24$, $30$, and $36$. Based on the full-scale simulations, we numerically identify the critical energy barrier $E_c^{nu}$, the critical impact amplitude $A_c$, and the critical total input energy $E_c^{in}$ for the three cases, which are reported in Fig.~\ref{fig:Collision_0deg_mode1}(g). 
	As expected, the critical impact amplitude $A_c$ and total input energy $E_c^{in}$ increase as the diameter increases, because 
	the nonlinear pulse spreads in the 2D domain, and therefore its amplitude spatially decays as it propagates through the media. In contrast, the critical energy barrier $E_c^{nu}$ shows no statistically significant change 
	(the small differences may be caused by inevitable numerical errors). 
	
\begin{figure*}[htbp]
		\centerline{ \includegraphics[width=1\textwidth]{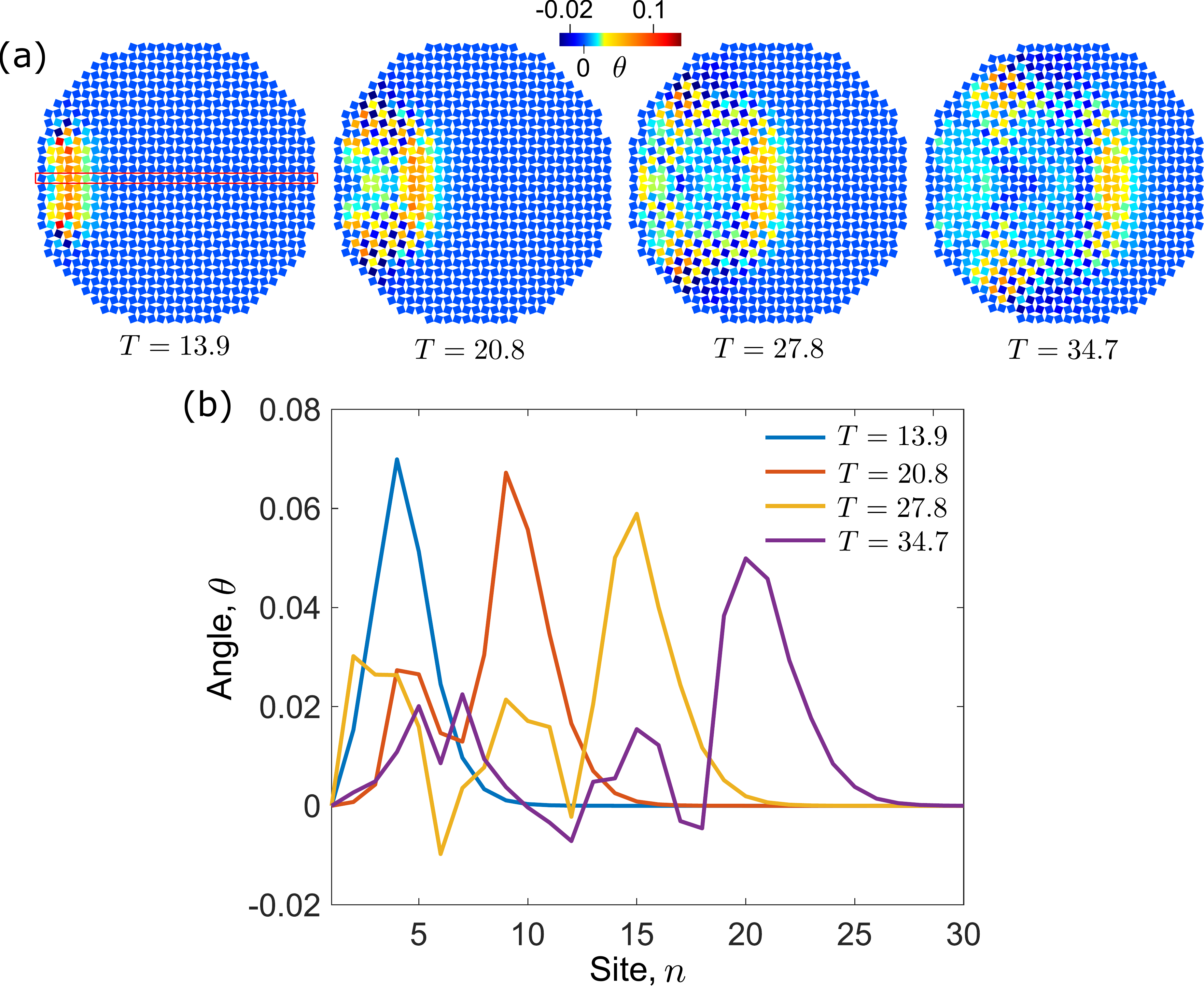}}
		\caption{Spreading of 
			a single pulse. (a) Snapshots of wavefields for a single pulse at $T=13.9$, $20.8$, $27.8$, and $34.7$. (b) Spatial profiles of the pulse along the propagation direction highlighted by the red lines.} 
		\label{fig:Dispersion_effects}
	\end{figure*}

	To further investigate the spreading of the pulses 
	mentioned above, we consider the propagation of a single pulse. Fig.~\ref{fig:Dispersion_effects}(a) shows snapshots from the numerical simulation of a single pulse propagation at normalized times $T=13.9$, $20.8$, $27.8$, and $34.7$, and the corresponding spatial profiles of the pulse along its propagation direction are given in Fig.~\ref{fig:Dispersion_effects}(b). 
	We observe dispersion, especially in the direction perpendicular to propagation, which is qualitatively similar to the expected 2D dispersion behavior observed previously \cite{Deng2019}. As a result, the amplitude of the pulse decreases as it propagates through the media.
	
	\subsection*{4.6 Characterization of the anisotropic behavior of the nonlinear pulses}
\begin{figure*}[htbp]
		\centerline{ \includegraphics[width=1\textwidth]{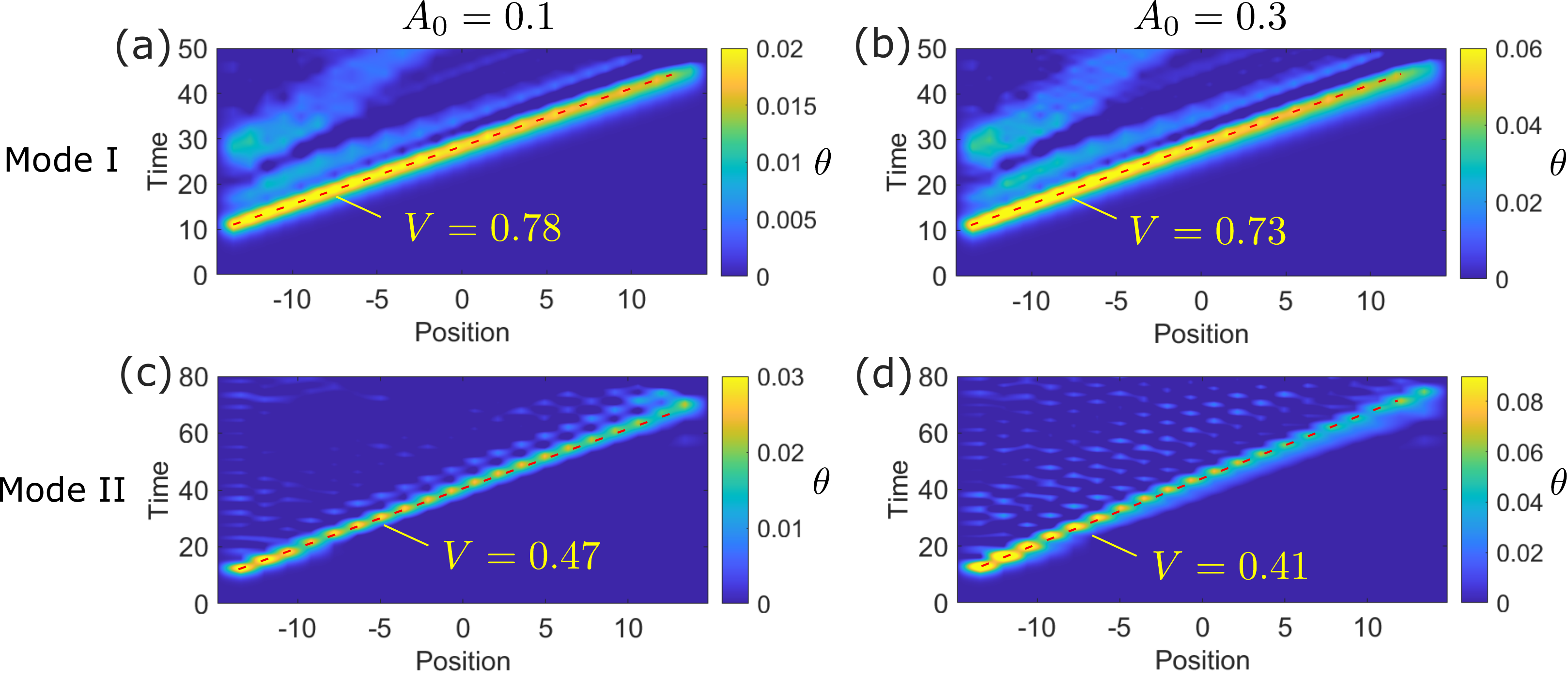}}
		\caption{Spatio-temporal plots extracted along the propagation direction for (a) pulse mode I with $A_0=0.1$; (b) pulse mode I with $A_0=0.3$; (c) pulse mode II with $A_0=0.1$; (b) pulse mode II with $A_0=0.3$.} 
		\label{fig:Contour_plot_SI}
	\end{figure*}

\begin{figure*}[htbp]
		\centerline{ \includegraphics[width=1\textwidth]{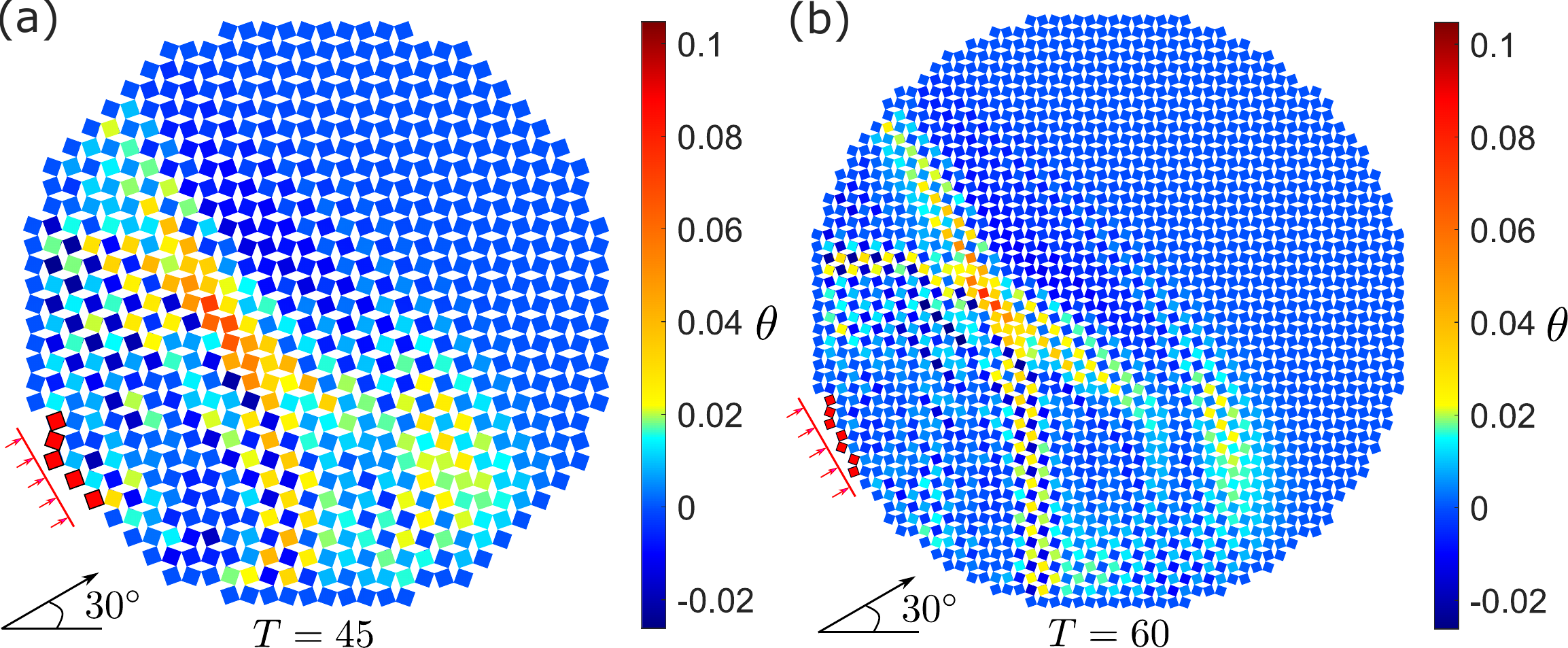}}
		\caption{Snapshots of wavefields with impact angle of $30^{\circ}$ for (a) a circular system with 30 squares in diameter at $T=45$ and (b) a circular system with 50 squares in diameter at $T=60$. The red squares in (a) and (b) are those to which the impact is applied.}
		\label{fig:Wavefield_30deg_SI}
	\end{figure*}

	Similar to the previous discussion of transition waves in \textbf{4.1}, we show in Fig.~\ref{fig:Contour_plot_SI} the contour plots for mode-I and mode-II pulses using the spatiotemporal data of the angles for two different impact amplitudes ($A_0=0.1$ and $0.3$). From these contour plots, we can approximately calculate the wave speed for each case, as reported in Fig.~\ref{fig:Contour_plot_SI}. 
	The wave speed of mode I is much faster than that of mode II. Moreover, the wave speeds associated with 
	both modes slightly decrease as the impact amplitude increases from $A_0=0.1$ to $0.3$. Moreover, we reported in Fig.~\ref{fig:Wavefield_30deg_SI} the snapshots for impact angle of $30^{\circ}$. We find that the wave separate into two modes with different wave speeds. Comaparing Fig.~\ref{fig:Wavefield_30deg_SI}(a) and (b), we observe that this separation behavior is more pronounced in a larger structure. 
	These findings are consistent with previous work on a monostable system of rotating squares~\cite{Deng2019}.  
	
	\clearpage
	
	\textbf{Movie 1}: Experimental observation of a phase transition in a 2D multistable metamatrial consisting of $10\times10$ rotating squares. A quasistatic load is applied at the center of the structure to trigger the phase transition.  The phase transition propagates outward throughout the rest of the structure in the form of a transition wave, transforming it from its initial open state (Phase~\textit{C}) to a closed state (Phase~\textit{R}).\\
	
	\textbf{Movie 2}: Numerical simulation of a phase transition induced quasistatically at the center of a structure comprising $30\times30$ squares, showing qualitative agreement with the experimental observations. \\
	
	\textbf{Movie 3}: Nucleation of a phase transition 
	via a head-on collision of soliton-like pulses. Two pulses with the same (positive) rotational direction are triggered by two 
	compressive impulses of amplitude $A_0=0.306\equiv A_c$ at the left and right boundary of a circular-shaped system. When the two pulses collide at the center, a critical nucleus of $2\times2$ squares of Phase~\textit{R} is formed. Then, the new phase propagates outward via transition waves.\\
	
	\textbf{Movie 4}:A head-on collision of two pulses with the same (positive) rotational direction for impact amplitude $A_0=0.3< A_c$. The critical nucleus is not formed and no phase transition is observed.\\
	
	\textbf{Movie 5}: Control of the location of nucleation via the timing of the impulses. The simulation on the left is obtained with impulses initiated at $\Delta T=10$, while the simulation on the right is obtained with impulses initiated at $\Delta T=20$. $\Delta T$ denotes the time delay of the impact applied at the left boundary with respect to the other impact.\\
	
	\textbf{Movie 6}: A head-on collision of two pulses with opposite rotational directions, with impact amplitude $A_0=A_c$. The two pulses pass through each other without nucleating a transition.\\
	
	\textbf{Movie 7}: Effects of propagation direction on the ability of colliding pulses to nucleate the new phase. Collision scenarios include (appearing in order): 1.~Collision of two mode-I pulses propagating along $x$ and $y$, with 
	$A_0=0.292$;  2.~Head-on collision of two mode-II pulses along the diagonal for $A_0=0.278$; 3. Collision of two mode-II pulses propagating 
	perpendicularly, with $A_0=0.24$; 4. Collision of a mode-I pulse and a mode-II pulse propagating along directions oriented $135$ degrees with respect to one another, with $A_0=0.302$; 5. Collision of a mode-I pulse and a mode-II pulse propagating along directions oriented $45$ degrees with respect to one another, with $A_0=0.314$.  
	
	\newpage
	
\end{document}